\documentclass[twocolumn]{aastex701}
\usepackage{float}
\usepackage{placeins}

\newcommand{\msun}{M$_{\odot}$}

\newcommand{\ms}{m\,s$^{-1}$}

\newcommand{\umich}{University of Michigan, Department of Astronomy, 1085 South University, Ann Arbor MI, 48109, USA}
\newcommand{\JPL}{Jet Propulsion Laboratory, California Institute of Technology, 4800 Oak Grove Dr., Pasadena, CA 91109, USA}
\newcommand{\nexsci}{NASA Exoplanet Science Institute, Caltech/IPAC, 1200 E. California Blvd., MC 100-22, Pasadena, CA 91125, USA}
\newcommand{\carnegieEPL}{Earth and Planets Laboratory, Carnegie Institution for Science, 5241 Broad Branch Road, NW, Washington, DC 20015, USA}
\newcommand{\carnegie}{The Observatories of the Carnegie Institution for Science, 813 Santa Barbara St., Pasadena, CA 91101, USA}
\newcommand{\UCR}{Department of Earth and Planetary Sciences, University of California, Riverside, CA 92521, USA}
\newcommand{\UCLA}{Department of Physics \& Astronomy, University of California, Los Angeles, Los Angeles, CA 90095, USA}
\newcommand{\UCIrvine}{Department of Physics \& Astronomy, The University of California, Irvine, Irvine, CA 92697, USA}
\newcommand{\UCBerkeley}{Department of Astronomy, University of California, Berkeley, CA 94720, USA}
\newcommand{\UCObservatories}{UC Observatories, University of California, Santa Cruz, CA 95064, USA}
\newcommand{\McDonald}{McDonald Observatory and Department of Astronomy, The University of Texas at Austin, TX 79734, USA}
\newcommand{\UTAustinPlanet}{Center for Planetary Systems Habitability, The University of Texas at Austin, 2305 Speedway Stop C1160, Austin, TX 78712-1692, USA}
\newcommand{\UTAustinAstro}{Department of Astronomy, The University of Texas at Austin, TX 78712, USA}
\newcommand{\UWMadison}{Department of Astronomy, University of Wisconsin-Madison, 475 N Charter St, Madison, WI 53706, USA}
\newcommand{\AMES}{NASA Ames Research Center, Moffett Field, CA 94035, USA}
\newcommand{\USydney}{Sydney Institute for Astronomy, School of Physics, The University of Sydney, Sydney, 2006, NSW, Australia}
\newcommand{\UA}{Department of Astronomy and Steward Observatory, University of Arizona, Tucson, AZ 85721, USA}

\newcommand{\NMSU}{Department of Astronomy, New Mexico State University, P.O. Box 30001, MSC 4500, Las Cruces, NM 88003, USA}
\newcommand{\LCO}{Las Campanas Observatory, Carnegie Institution for Science Colina el Pino, Casilla 601 La Serena, Chile}
 
\begin{document}

\title{A Century of Radial Velocity and Astrometric Monitoring of 70 Oph AB: New PFS Data and Constraints on \textcolor{black}{Possible} Planetary Companions  \footnote{This paper includes data gathered with the 6.5 meter Magellan Telescopes located at Las Campanas Observatory, Chile.}}

\author[0000-0002-6845-9702]{Yiting Li}
\affiliation{\umich}
\email{lyiting@umich.edu}

\author[0000-0003-1227-3084]{Michael R. Meyer}
\affiliation{\umich}
\email{mrmeyer@umich.edu}

\author[0000-0001-5592-6220]{Skylar D'Angiolillo}
\affiliation{\UCR}
\email{sdang049@ucr.edu}

\author[0000-0002-7084-0529]{Stephen R. Kane}
\affiliation{\UCR}
\email{skane@ucr.edu}

\author[0000-0003-1305-3761]{R. Paul Butler}
\affiliation{\carnegieEPL}
\email{pbutler@carnegiescience.edu}

\author[0000-0002-8681-6136]{Stephen A. Shectman}
\affiliation{\carnegie}
\email{shec@carnegiescience.edu}

\author[0000-0003-2008-1488]{Eric E. Mamajek}
\affiliation{\JPL}
\email{eric.mamajek@jpl.nasa.gov}

\author[0009-0008-2801-5040]{Johanna Teske}
\affiliation{\carnegieEPL}
\affiliation{\carnegie}
\email{jteske@carnegiescience.edu}

\author[0000-0001-8342-7736]{Jack Lubin}
\affiliation{\UCLA}
\email{jblubin@astro.ucla.edu}

\author[0000-0003-0149-9678]{Paul Robertson}
\affiliation{\UCIrvine}
\email{paul.robertson@uci.edu}

\author[0000-0002-8035-4778]{Jessie L. Christiansen}
\affiliation{\nexsci}
\email{christia@ipac.caltech.edu}

\author[0000-0002-0531-1073]{Howard Isaacson}
\affiliation{\UCBerkeley}
\email{hisaacson@berkeley.edu}

\author[0000-0001-5737-1687]{Caleb K. Harada}
\affiliation{Institute for Astronomy, University of Hawai`i, 2680 Woodlawn Drive, Honolulu, HI 96822, USA}
\affiliation{Department of Earth Sciences, University of Hawai`i at Man\=oa, 1680 East-West Road, Honolulu, HI 96822, USA}
\email{charada@berkeley.edu}

\author[0000-0002-6153-3076]{Bradford Holden}
\affiliation{\UCObservatories}
\email{holden@ucolick.org}

\author[0000-0001-9662-3496]{William D. Cochran}
\affiliation{\McDonald}
\affiliation{\UTAustinPlanet}
\email{wdc@astro.as.utexas.edu}

\author[0000-0002-7714-6310]{Michael Endl}
\affiliation{\UTAustinPlanet}
\affiliation{\UTAustinAstro}
\email{mike@astro.as.utexas.edu}

\author[0000-0002-0040-6815]{Jennifer Burt} 
\affiliation{\JPL}
\email{}

\author[0000-0002-7733-4522]{Juliette Becker}
\affiliation{\UWMadison}
\email{juliette.becker@wisc.edu}

\author[0000-0003-4287-004X]{Alyssa Jankowski}
\affiliation{\UWMadison}
\email{alyssajanko5@gmail.com}

\author[0000-0001-7026-6291]{Peter Tuthill}
\affiliation{\USydney}
\email{peter.tuthill@sydney.edu.au}

\author[0000-0002-2361-5812]{Catherine A. Clark}
\affiliation{\nexsci}
\email{clarkc@ipac.caltech.edu}

\author[0000-0002-9288-3482]{Rachael M.\ Roettenbacher}
\affiliation{\umich}
\email{}

\author[0000-0001-6975-9056]{Eric Nielsen}
\affiliation{\NMSU}
\email{nielsen@nmsu.edu}

\author[0000-0002-9408-8925]{Eduardo Bendek}
\affiliation{\AMES}
\email{eduardo.bendek@nasa.gov}

\author[0000-0002-4675-9069]{Armen Tokadjian}
\affiliation{\JPL}
\email{armen.tokadjian@jpl.nasa.gov}

\author[0000-0000-0000-0000]{William Roberson}
\affiliation{\NMSU}
\email{wcr@nmsu.edu}

\author[0000-0001-5253-1338]{Kaitlin M. Kratter}
\affiliation{\UA}
\email{kkratter@arizona.edu}

\author[0000-0003-4179-6394]{Edwin Bergin}
\affiliation{\umich}
\email{}

\author[0000-0003-0412-9664]{Dave Osip}
\affiliation{\LCO}
\email{dosip@carnegiescience.edu}

\author[0000-0002-5226-787X]{Jeffrey D. Crane}
\affiliation{\carnegie}
\email{crane@carnegiescience.edu}

\author[0000-0003-2082-5176]{Alex Davis}
\affiliation{\JPL}
\email{alex.davis@jpl.nasa.gov}

\author[0000-0002-1871-6264]{Gautam Vasisht}
\affiliation{\JPL}
\email{gautam.vasisht@jpl.nasa.gov}

\begin{abstract}

At a distance of 5.1~pc, the 70~Oph~AB binary star system is one of the most favorable targets for future direct imaging and astrometry missions surveying mature, terrestrial planets. We present new radial velocities (RVs) obtained with the Planet Finder Spectrograph (PFS) on the 6.5\,m Magellan~II Clay Telescope in Chile. We collected 499 measurements of 70~Oph~A and 334 measurements of 70~Oph~B during 2023--2025. Combining these data with decades of archival RVs and astrometry, we derive an updated orbital solution for the binary and dynamical masses of $0.88 \pm 0.004\,M_\odot$ and $0.73 \pm 0.003\,M_\odot$ for the primary and secondary components, respectively. We find that the long-term RV variability of both components is consistent with stellar activity modulated by rotation periods, and we detect no coherent planetary signals in either component. We place upper limits on any planets orbit in the plane of the binary. The 27-yr RV baseline for 70~Oph~A excludes Jupiter-mass planets interior to 5 au and reaches sensitivity of $0.3\,M_{\rm Jup}$ at 1 au or $0.5\,M_{\rm Jup}$ at 2 au. For 70~Oph~B, with PFS data we rule out planets more massive than 0.25-0.3$\,M_{\rm Jup}$ inside 0.5~AU. We show that stable S-type orbits around 70~Oph~A extend to $\sim2.5$~AU, covering the habitable zone. Thus, Saturn-mass planets or smaller on stable orbits in the habitable zone of 70~Oph~A are allowed. Overall, our results provide important guidance for future planet searches around this stellar system.

\end{abstract}



\section{Introduction}
\label{sec:intro}

Over the past two decades, the number of confirmed exoplanets has grown to more than 6,000.\footnote[1]{From the NASA Exoplanet Archive:  \url{https://exoplanetarchive.ipac.caltech.edu/}.} This remarkable progress has been driven primarily by transit and radial-velocity (RV) surveys, which have steadily lowered detection thresholds toward terrestrial-mass planets. Space missions including \textit{CoRoT} \citep{Baglin_2006}, \textit{Kepler} \citep{Borucki2010}, \textit{TESS} \citep{Ricker2015}, and \textit{CHEOPS} \citep{Benz2021} have identified thousands of transiting systems. However, transit surveys are intrinsically limited to favorable orbital geometries. The detection and characterization of non-transiting planets, particularly in the habitable zones (HZs) of nearby stars, currently rely on high-precision, ground-based spectroscopy. Thanks to advances in RV instrumentation, data processing, and observing strategies, high-precision spectrographs such as HARPS-N \citep{Cosentino2012}, CARMENES \citep{Quirrenbach2014}, HPF \citep{Mahadevan2012}, PFS \citep{Crane2006,Crane2008,Crane2010}, and newer facilities such as NEID \citep{Schwab_2016}, ESPRESSO \citep{Pepe_2013}, and KPF \citep{Gibson_2024} are able to achieve sub-m/s level precision, enabling the detection of super-Earths and sub-Neptunes around nearby, slowly rotating stars.

RV surveys of single FGK stars have established strong correlations between giant planet occurrence and stellar metallicity \citep{Fischer2005, Santos2004}, as well as stellar mass \citep{Johnson2010, Bowler2021}. However, nearly half of solar-type stars reside in binary or multiple systems \citep{Raghavan2010}, and the planet occurrence rate in these systems is still poorly understood because the presence of a stellar companion can complicate both planet formation and long-term orbital stability \citep{Hirsch_2021}. Recent studies show that planet occurrence is suppressed by a factor of $\sim$2--3 in binaries with separations $\lesssim$50~AU \citep{Wang2014, Kraus2016, Moe_2021}, likely because tidal interactions truncate circumstellar disks, which depletes the reservoir of material available for planet formation. Long-term orbital stability is further challenged by secular perturbations, such as Kozai--Lidov cycles, which can drive large oscillations in eccentricity and inclination \citep{Kozai1962, Lidov1962, MurrayDermott1999}. And yet, the detection of S-type planets (orbiting a single component within a binary) in systems such as Proxima~b \citep{Anglada2016}, $\gamma$~Cephei, and HD~196885 \citep{Hatzes2003, Correia2008} demonstrates that planet formation can occur in complex binary environments. Thus far, about 75\% of S-type planets in binaries have been detected via the transit method, with most of the rest via the RV technique \citep{Thebault2025}. More recently, direct-imaging searches have also begun to probe this parameter space. Observations of $\alpha$~Centauri~A \citep{Beichman_2025,Sanghi_2025, Bendek_2026}, for example, reported candidate giant planets in the system and placed stringent constraints on planets at small separations \cite{Wagner_2021}.

Complementary to these efforts, astrometry is another technique that is particularly powerful for long-period or low-inclination systems, because joint RV and astrometric measurements can constrain orbital inclination and thus reveal the true companion mass \citep{Brandt2018, Li_2021}. Whereas RV measures the line-of-sight component of stellar reflex motion, absolute astrometry from \textit{Hipparcos} and \textit{Gaia}, in addition to relative astrometry from direct imaging, captures the orthogonal plane-of-sky component. Since the \textit{Hipparcos} mission \citep{Perryman1997}, astrometric precision has improved by nearly two orders of magnitude with \textit{Gaia} \citep{Gaia2016, Gaia2023}, reaching $\sim$10--100~$\mu$as. Combined with long-baseline RV data \citep[e.g.,][]{Brandt2018,Brandt2021}, these measurements yield well-constrained orbital solutions and robust dynamical mass determinations.

Looking ahead, next-generation direct-imaging and high-precision astrometric missions will further expand the capability to detect habitable planets in binary systems. On the ground, 30-m class Extremely Large Telescopes (ELTs) equipped with instruments such as the Mid-infrared ELT Imager and Spectrograph (METIS) \citep{Brandl_2010} will detect planets at small angular separations through thermal emission at mid-infrared wavelengths \citep{Bowens_2021}. In space, the Habitable Worlds Observatory (HWO; \citealt{Gaudi_2020,Feinberg,Harada_2024}) is expected to directly image nearby potentially habitable planets in reflected light, while proposed future instruments such as the {\it TOLIMAN} mission \citealt{Toliman2018} and the {\it Searching for Habitable Exoplanets with Relative Astrometry} (SHERA; \citealt{Christiansen2025SHERA}) mission that aim to  provide precise astrometric measurements for a select sample of the nearest stars, including binary systems such as $\alpha$~Cen~AB, 36~Oph~AB, and 70~Oph~AB \citep{Mamajek_2024, Christiansen2025SHERA}. Therefore, establishing precise orbital solutions for these systems is essential for target prioritization and yield analyses.

Among the nearest binary star systems, 70~Oph~AB (HD~165341~AB) is famous for being the subject of one of the earliest exoplanet claims, made by \citet{Jacob1855}. However, that putative planet was quickly discarded as an artifact of nineteenth-century visual micrometry \citep{Campbell1988,Walker1995}, and no planets have been reported in the system. Nevertheless, the system’s century-long radial velocity record now provides unique sensitivity for a comprehensive planet search. In this paper, we present new high-precision RVs from the Planet Finder Spectrograph (PFS), the most precise dataset obtained for this system to date. Leveraging archival RVs and \textit{Hipparcos}-\textit{Gaia} absolute astrometry, we refine the orbital solution for 70~Oph~AB and derive upper limits on potential planets in the system.

This paper is organized as follows. 
Section~\ref{sec:stellarproperty} summarizes the stellar properties of 70~Oph~AB, including activity and rotation diagnostics. Section~\ref{sec:observations} describes our new PFS observations and data reduction. Section~\ref{sec:orbit} presents the joint radial-velocity and astrometric modeling of the binary orbit and the resulting dynamical masses. Section~\ref{sec:companionsearch} describes the search for planetary companions using periodogram analysis. In Section~\ref{sec:GPR}, we model the RV variability with a series of Gaussian-process regression models. We then quantify detection limits through injection–recovery tests in Section~\ref{sec:injection-recovery} and assess dynamical stability constraints on potential S-type planets using dynamical simulations in Section~\ref{sec:limits}. 
Section~\ref{sec:discussions} discusses the implications of our results, and Section~\ref{sec:conclusions} summarizes our conclusions.

\begin{table}
\centering
\scriptsize
\caption{Stellar Properties for 70~Oph AB \label{tab:stellar_params}}
\begin{tabular}{lcc}
\hline\hline
Parameter & 70~Oph~A & 70~Oph~B \\
\hline
\multicolumn{3}{c}{\textbf{Main Identifiers}} \\
HD & 165341~A & 165341~B \\
GJ & 702~A & 702~B \\
WDS & J18055+0230A & J18055+0230B \\
Gaia~DR3 & 4468557611984384512 & 4468557611977674496 \\
\hline
\multicolumn{3}{c}{\textbf{Astrometry (Gaia DR3)}} \\
$\alpha_{\rm ICRS}$ (deg) & 271.3645 & 271.3659 \\
$\delta_{\rm ICRS}$ (deg) & $+2.4952$ & $+2.4942$ \\
Parallax (mas) & $195.5674 \pm 0.1964$ & $195.8563 \pm 0.2535$ \\
Distance $d$ (pc) & $5.111 \pm 0.005$ & $5.105 \pm 0.007$ \\
$\mu_\alpha$ (mas\,yr$^{-1}$) & $+206.53$ & $+333.29$ \\
$\mu_\delta$ (mas\,yr$^{-1}$) & $-1107.49$ & $-1068.35$ \\
RUWE & 3.27 & 3.71 \\
\hline
\multicolumn{3}{c}{\textbf{Stellar Properties}} \\
Spectral type & K0\,V & K4\,V \\
$T_{\rm eff}$ (K) & $5301 \pm 67$ & $4465 \pm 65$ \\
$\log g$ (cgs) & $4.55 \pm 0.03$ & $4.64 \pm 0.02$ \\ 
$[{\rm Fe/H}]$ (dex) & $+0.05 \pm 0.03$ & $+0.05 \pm 0.03$ \\
$R_\star$ ($R_\odot$) & $0.831 \pm 0.004$ & $0.670 \pm 0.009$ \\
$L_\star$ ($L_\odot$) & $0.53 \pm 0.02$ & $0.15 \pm 0.02$ \\
\hline
\multicolumn{3}{c}{\textbf{Activity and Rotation}} \\
$\log R'_{\rm HK}$ & $-4.565 \pm 0.010$ & $-4.710 \pm 0.015$ \\
$P_{\rm rot}$ (days) & $19.33 \pm 0.31$ & $24.8 \pm 1.0$ \\
$\log(L_X/L_{\rm bol})$ & $-4.95$ & $-5.17$ \\
\hline
\multicolumn{3}{c}{\textbf{Masses and Age}} \\
$M_\star$ ($M_\odot$)$^{\dagger}$ & $0.89 \pm 0.02$ & $0.73 \pm 0.01$ \\
$M_\star$ ($M_\odot$)$^{\ast}$    & $0.8827^{+0.0044}_{-0.0043}$ & $0.7319^{+0.0031}_{-0.0031}$ \\
Age (Gyr)$^{\ddagger}$ & $2.81^{+1.65}_{-0.80}$ & $2.99^{+0.66}_{-0.51}$
\\
Age (Gyr)$^{\S}$ & $6.2 \pm 1.0$ & $6.2 \pm 1.0$ \\
\hline
\end{tabular}

\vspace{1mm}
\raggedright
\footnotesize
$^{\dagger}$ Literature masses from \citet{Eggenberger2008}. \\
$^{\ast}$ Masses from this work (Section~\ref{sec:orbit}). \\
$^{\ddagger}$ Activity-based ages from this work. \\
$^{\S}$ Asteroseismic age of 70~Oph~A from \citet{Eggenberger2008}. Astrometry is from Gaia~DR3. Spectral classifications are from \citet{Cowley1967} and \citet{Gray2003}. Stellar atmospheric parameters ($T_{\rm eff}$, $\log g$, [Fe/H]) are adopted from \citet{Yee2017}. Stellar radii and luminosities are taken from interferometric measurements by \citet{Boyajian2012}. Chromospheric activity indices ($\log R'_{\rm HK}$) and rotation periods are from \citet{Olspert2018}. X-ray activity levels ($\log L_X/L_{\rm bol}$) are adopted from \citet{Ayres2022}.
\end{table}

\section{Stellar Properties}
\label{sec:stellarproperty}

\subsection{Overview}

70~Ophiuchi~AB is a nearby ($d = 5.1$~pc; \citealt{GaiaDR3}), well-studied visual and spectroscopic binary composed of two early K dwarfs. The system has been monitored for over a century through astrometric and radial-velocity observations, yielding an $\sim$88-year orbit. Extensive spectroscopic and interferometric studies provide well-constrained fundamental parameters for both stars. A summary of the stellar parameters for 70~Oph~AB is listed in Table~\ref{tab:stellar_params}.

\subsection{Fundamental Stellar Parameters}

The spectral classifications of 70~Oph~AB are K0\,V (A) and K4--K5\,V (B), respectively \citep{Cowley1967,Gray2003}. Spectroscopic analyses of high-resolution HARPS data \citep{Carrier2006}, together with subsequent literature studies \citep[e.g.,][]{Houdebine2019,Yee2017,Soubiran2022}, yield an effective temperature of $T_{\rm eff} = 5300 \pm 50$~K for 70~Oph~A. For the secondary, we adopt $T_{\rm eff,B} = 4465 \pm 65$~K from the spectroscopic analysis of \citet{Yee2017}.

Both stars in 70~Oph exhibit near-solar metallicity, with a mean value of $[{\rm Fe/H}] \approx +0.04 \pm 0.05$ and no evidence for $\alpha$-element enhancement \citep{Santos2004}. Interferometric radius measurements yield $R_A = 0.83 \pm 0.01\,R_\odot$ and $R_B = 0.67 \pm 0.01\,R_\odot$ \citep{Boyajian2012,Eggenberger2008}, corresponding to luminosities of $L_A = 0.53 \pm 0.02\,L_\odot$ and $L_B = 0.15 \pm 0.02\,L_\odot$. These radii and luminosities indicate that both stars are unevolved main-sequence K dwarfs of near-solar metallicity.

\subsection{Rotation Periods}

Rotation periods for 70~Oph~A and B have been inferred from long-term chromospheric and coronal activity diagnostics. Mount Wilson $S$-index monitoring over multiple decades provides rotation estimates for both components \citep{Baliunas1996}. For 70~Oph~A, measurements consistently yield $P_{\rm rot} \approx 19$--20~days, including $P_{\rm rot} = 19.33 \pm 0.31$~days from \citet{Olspert2018}, $19.5 \pm 0.3$~days from \citet{Ayres2022}, and a mean value of 20~days reported by \citet{Baliunas1996}. In contrast, the rotation period of 70~Oph~B is less constrained. Mount Wilson measurements suggest $\langle P_{\rm rot} \rangle \approx 34$~days \citep{Baliunas1996}, while activity-based calibrations and X-ray studies indicate a shorter range of $\sim$23--25~days \citep[e.g.,][]{Noyes1984,Ayres2022}. 

We independently analyzed a Generalized Lomb--Scargle (GLS) periodogram \citep{Zechmeister_2009} of the archival Mount Wilson Ca\,II H\&K $S$-index data \citep{Duncan_1991,MWO_HK_Data}. After removing a quadratic trend to account for long-term variability, we recover a strong 20-day peak consistent with the published rotation period of 70~Oph~A. Our window function is dominated by a 365-day annual alias and shows negligible power near 20~days and elsewhere. Unfortunately, the available Mount Wilson $S$-index data for 70~Oph~B are insufficient to constrain its rotation period.

Public \textit{TESS} photometry \citep{Ricker2015} is available for the unresolved 70~Oph~AB system (TIC~398120047) from the Mikulski Archive for Space Telescopes (MAST) \footnote[2]{ \url{http://dx.doi.org/10.17909/285z-xh42}.}, but only for a single $\sim$27-day Sector 80 light curve. This light curve does not provide a meaningful constraint on the rotation period because the observing baseline is too short, the photometry is blended, and the standard TESS pipeline tends to detrend astrophysical signals longer than about 10 days.

%Public \textit{TESS} photometry \citep{Ricker2015} is also available for 70~Oph~AB (TIC~398120047) from the Mikulski Archive for Space Telescopes (MAST), obtained during the $\sim$27-day Sector 80. Because the binary separation ($\sim6.8\arcsec$) is much smaller than the \textit{TESS} pixel scale ($21\arcsec$), the light curve contains the combined flux of both components. We plot the detrended TESS PDCSAP light curve in Figure~\ref{fig:tess}. We find that the light curve exhibits low-amplitude, quasi-periodic variability on $\sim$20-day timescales, and the corresponding GLS periodogram shows a peak at $P = 20.64$~days. This periodicity is consistent with the 20-day rotation period of 70~Oph~A. However, the blended photometry does not allow us to distinguish between the rotation periods of the individual components.

\subsection{Age from Activity and Asteroseismology}

% \begin{figure}
% \centering
% \includegraphics[width=\linewidth]{70Oph_TESS_LC_Periodogram.pdf}
% \caption{
% Light curve from \textit{TESS} Sector~80 photometry of the unresolved 70~Oph~AB binary system (TIC~398120047). 
% The top panel shows the normalized light curve, exhibiting low-amplitude, quasi-periodic variability. The dotted dips are caused by instrumental or systematic events. The bottom panel displays the Generalized Lomb--Scargle periodogram over 5--30~days, with a peak at $P = 20.64$~days consistent with the literature rotation timescale of the system. Because the system is unresolved in the \textit{TESS} aperture, variability from the two components is blended, and the signal cannot be uniquely assigned to either star.}
% \label{fig:tess}
% \end{figure}

The age of 70~Oph has been estimated using rotation- and activity-based diagnostics as well as asteroseismology. We first determine gyrochronological ages by applying calibrated rotation-age relations \citep{Mamajek2008, Bouma2023} to the periods and effective temperatures listed in Table~\ref{tab:stellar_params}. This yields $2.21^{+0.21}_{-0.17}$~Gyr for 70~Oph~A and $3.16^{+0.37}_{-0.33}$~Gyr for 70~Oph~B.% Assuming coevality, the discrepancy of $\sim$2.4$\sigma$ likely arises from intrinsic scatter in the period-age relation or systematic uncertainties inherent to gyrochronology across different spectral types.

To further constrain the system age, we employ an activity-based approach following the method outlined by \citet{Li_2021}, which utilizes the Bayesian technique of \citet{Brandt_2014} to derive age posteriors from multiple magnetic activity proxies. This method incorporates chromospheric Ca~\textsc{ii}~H\&K emission ($\log R'_{HK}$) from the Mt. Wilson $S$-index survey, coronal X-ray luminosities ($\log L_X/L_{\rm bol}$) observed by the \textit{ROSAT} satellite (e.g., \citealt{Ayres2022}), and observed rotation periods ($P_{\rm rot}$) to calculate the Rossby number based on the empirical relations of \citet{Mamajek2008}. We derive a median activity-based age of $2.81^{+1.65}_{-0.80}$~Gyr for 70~Oph~A and $2.99^{+0.66}_{-0.51}$~Gyr for 70~Oph~B. The agreement between the stellar ages resolves the tension found in the pure gyrochronological estimates. 

Despite finding an age of $\sim$3~Gyr from activity and gyrochronology, previous asteroseismic modeling of solar-like $p$-mode oscillations in 70~Oph~A suggested a substantially older age of $\sim$6$-$7~Gyr \citep{Carrier2006, Eggenberger2008}. One possible explanation for this discrepancy is weakened magnetic braking, which can explain similar disagreements in other solar-type stars. Studies of \textit{Kepler} asteroseismic targets suggest that once stars reach a critical Rossby number, changes in magnetic field topology can significantly reduce angular momentum loss, allowing them to retain relatively rapid rotation and elevated activity levels even at older ages \citep{vanSaders_2016, Metcalfe_2025}. In this case, activity-based age estimates may underestimate the true ages. An alternative possibility is that the published asteroseismic age estimate for 70~Oph~A, based on the modeling of \citet{Eggenberger2008}, is sensitive to the stellar input physics and abundance assumptions adopted at the time. We therefore caution that the inferred seismic age may be outdated and would benefit from reanalysis with modern stellar models. Ultimately, a modern astroseismic analysis with updated input physics could resolve the age discrepancy. 

%Despite finding a relatively young age of $\sim$3~Gyr based on activity and gyrochronology, asteroseismic modeling of solar-like $p$-mode oscillations in 70~Oph~A favors a substantially older age of $\sim$6--7~Gyr \citep{Carrier2006, Eggenberger2008}. This large discrepancy between activity-based and asteroseismic dating is a known phenomenon in solar-type stars and is likely explained by Weakened Magnetic Braking (WMB). Recent studies using \textit{Kepler} data demonstrate that once stars reach a critical Rossby number, they undergo a transition in magnetic field topology that leads to a dramatic reduction in angular momentum loss \citep{vanSaders_2016, Metcalfe_2022}. Consequently, an older star can maintain a youthful rotation period and high activity level, causing rotation- and activity-based models to underestimate the true chronological age. Given that both 70~Oph~A and B are consistent with the Rossby numbers where WMB is expected to occur, we consider the older asteroseismic age of $\sim$6~Gyr to be the more plausible absolute age for the system, while the $\sim$3~Gyr result may represent the gyrochronological age at which the system's magnetic evolution stalled.

\begin{figure}
\centering
\includegraphics[width=1\linewidth]{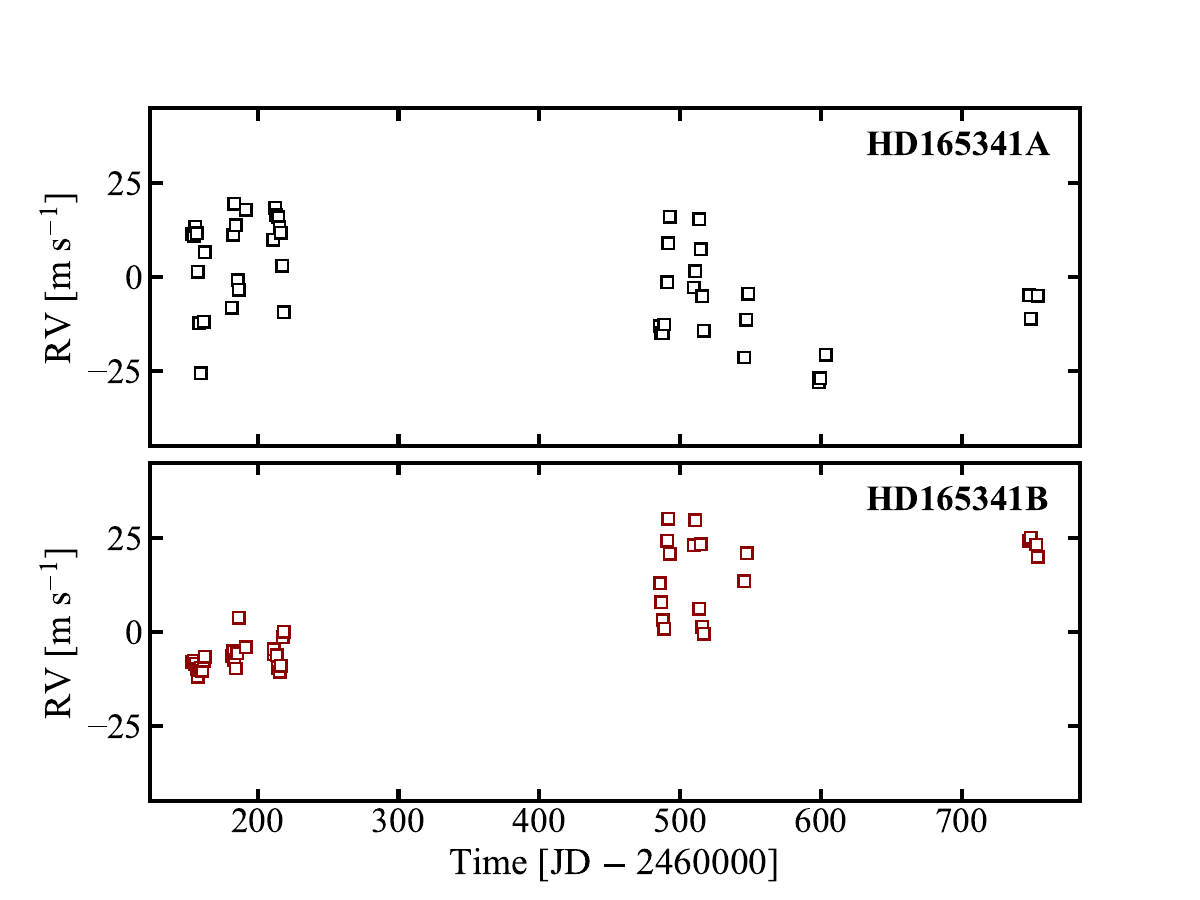}
\caption{
Nightly binned RVs of 70~Oph~A (48 nights) and 70~Oph~B (44 nights) derived from PFS spectra, shown as a function of time. Each night's data are represented by a single weighted-average measurement, with weights computed as the inverse square of the RV uncertainties. The time of each binned point corresponds to the weighted average of the observation times. Error bars are smaller than the symbol size and are therefore not visible.}
\label{fig:nightly_RVs}
\end{figure}

\begin{figure*}[!htb]
\centering
\hspace*{+11.6mm}
\includegraphics[width=0.425\linewidth]{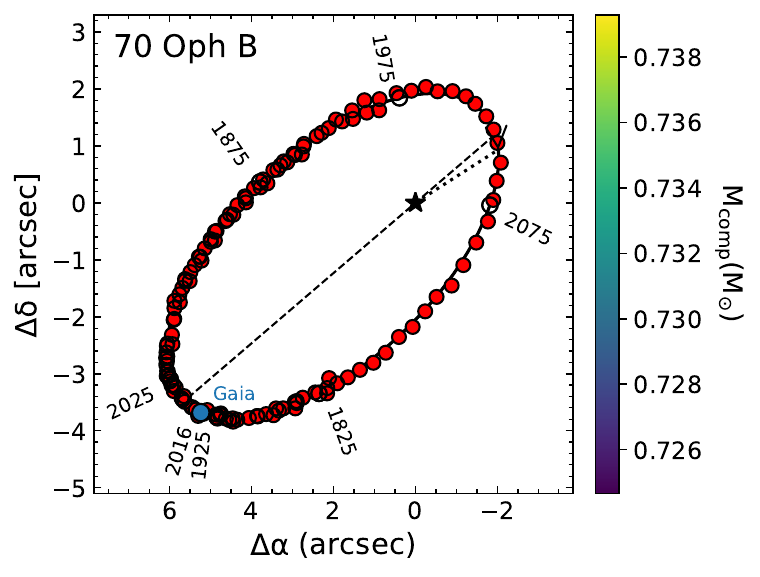}
\includegraphics[width=0.5\linewidth]{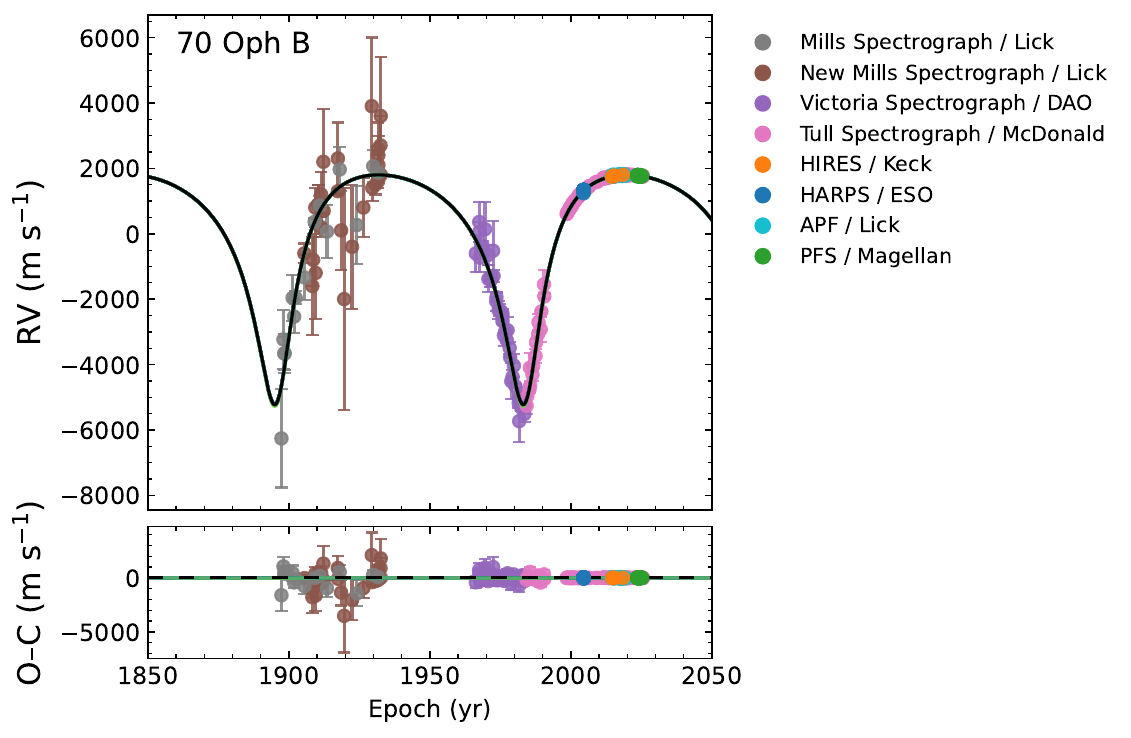}
\caption{The left panel shows the relative astrometric orbit of 70~Oph~B around 70~Oph~A. Red points denote archival visual and micrometric measurements from the WDS/ORB6 catalog, while the solid black curve shows the maximum-likelihood joint orbital solution. 50 orbits color-coded by companion mass are randomly drawn from the posterior and plotted, but they are indistinguishable from the black curve because of the tight constraint on the orbit. The dashed line marks the line of nodes, and the star indicates the primary's position. The right panel shows the RV variation of 70~Oph~A measured by eight instruments. Colors correspond to individual RV datasets as indicated in the legend. Residuals relative to the best-fit model are shown.  }
\label{fig:binary_orbit}
\end{figure*}

\begin{figure*}[!htb]
\centering
\hspace*{-2.2mm}
\includegraphics[width=0.4\linewidth]{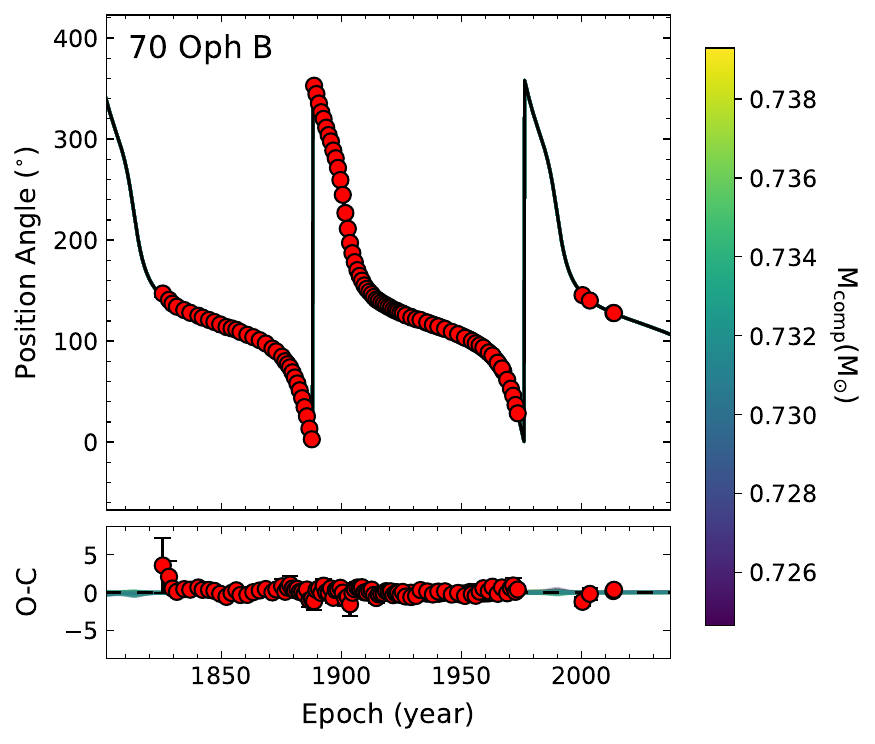}
\includegraphics[width=0.408\linewidth]{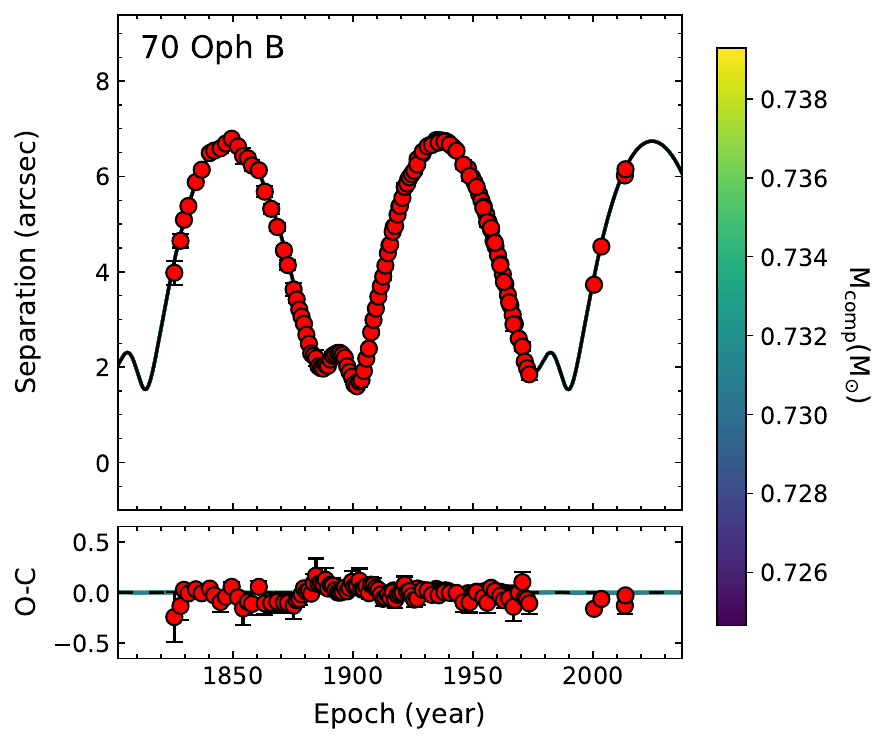}
\caption{The left panel displays the relative position angle vs. time, shown with the maximum-likelihood orbital solution from the joint astrometric and RV fit (black curve). Residuals relative to the best-fit model are shown. The right panel shows relative separation vs. time, again compared with the maximum-likelihood orbital model. 50 random orbits drawn from the posterior distribution are overplotted, which are colored by companion mass, and they are hidden behind the maximum-likelihood orbit.}
\label{fig:binary_orbit_2}
\end{figure*}

\section{Observations}\label{sec:observations}

We collected high-precision radial-velocity (RV) measurements of the 70~Ophiuchi~AB system (V = 4.03 and 4.98 for components A and B, respectively) with the Planet Finder Spectrograph \citep[PFS;][]{Crane2006, Crane2008, Crane2010} on the 6.5\,m Magellan Clay Telescope at Las Campanas Observatory as part of our long-term RV campaign targeting the nearest and brightest solar-type stars. Our observing campaign covered 48 nights between 2023~August and 2025~March, spanning 600 days. Figure~\ref{fig:nightly_RVs} shows the nightly binned RVs for each component, where individual exposures from a given night are combined into a single weighted-average measurement. \textcolor{black}{These radial velocity time series data are provided in full in the Appendix.}

All of our data were taken with the iodine-cell configuration of PFS and the STA1600 CCD, installed in 2017. This detector provides continuous wavelength coverage from 3880--6680\,\AA\ at a resolving power of $R \sim 127{,}000$ with a 0\farcs3\,$\times$\,2\farcs5 slit and 1$\times$2 binning. Exposure times were 600--900\,s. We typically recorded two or three exposures per night. These settings produced a median signal-to-noise ratio of about 230 per pixel at 5500\,\AA\ and internal RV uncertainties of 0.7--0.8\,m\,s$^{-1}$. We also obtained high-S/N template spectra for both stars with the same observing set-up sans iodine.

We processed all spectra and extracted RVs using the standard PFS iodine forward-modeling pipeline \citep{Butler1996}. Each observed spectrum is modeled as a Doppler-shifted stellar template multiplied by a high-resolution iodine transmission spectrum and convolved with a wavelength-dependent instrumental profile. This method routinely achieves $\sim$1\,m\,s$^{-1}$ precision on individual observations of bright stars. Instrumental stability was monitored using stable RV reference stars observed throughout the campaign. Long-term monitoring of reference stars indicates instrumental stability of approximately 0.5\,m\,s$^{-1}$ over the full observing baseline.

%We obtained the H$\alpha$ index from each iodine-free exposure to monitor stellar activity. 

Re-emission in the cores of the Ca~II H \& K and H$\alpha$ lines is a well-known proxy for stellar activity. We measured the H$\alpha$ index from each iodine exposure as a tracer of the chromospheric emission component in the H$\alpha$ Balmer line at 6563~\AA, following the definition of \citet{daSilva_2011}. Uncertainties in the H$\alpha$ index were estimated from photon noise in the continuum regions. The H$\alpha$ time series shows no long-term trends or discontinuities. Unfortunately, we were unable to obtain contemporaneous Ca~II H \& K measurements or other stellar activity indicators in addition to H$\alpha$.

%We measured the H$\alpha$ index from each RV exposure . Uncertainties for the H$\alpha$ index were estimated from photon noise in the continuum regions. The H$\alpha$ time series shows no long-term trends or discontinuities. Unfortunately, we were unable to obtain contemporaneous Ca II H$\&$K measurements or other stellar activity indicators besides H$\alpha$. The H$\alpha$ line was unaffected because it lies outside the iodine absorption region.

%Unfortunately, we were unable to obtain contemporaneous Ca II H$\&$K measurements or other stellar activity indicators besides H$\alpha$.

\begin{figure*}[!htb]
    \centering
    \includegraphics[width=\linewidth]{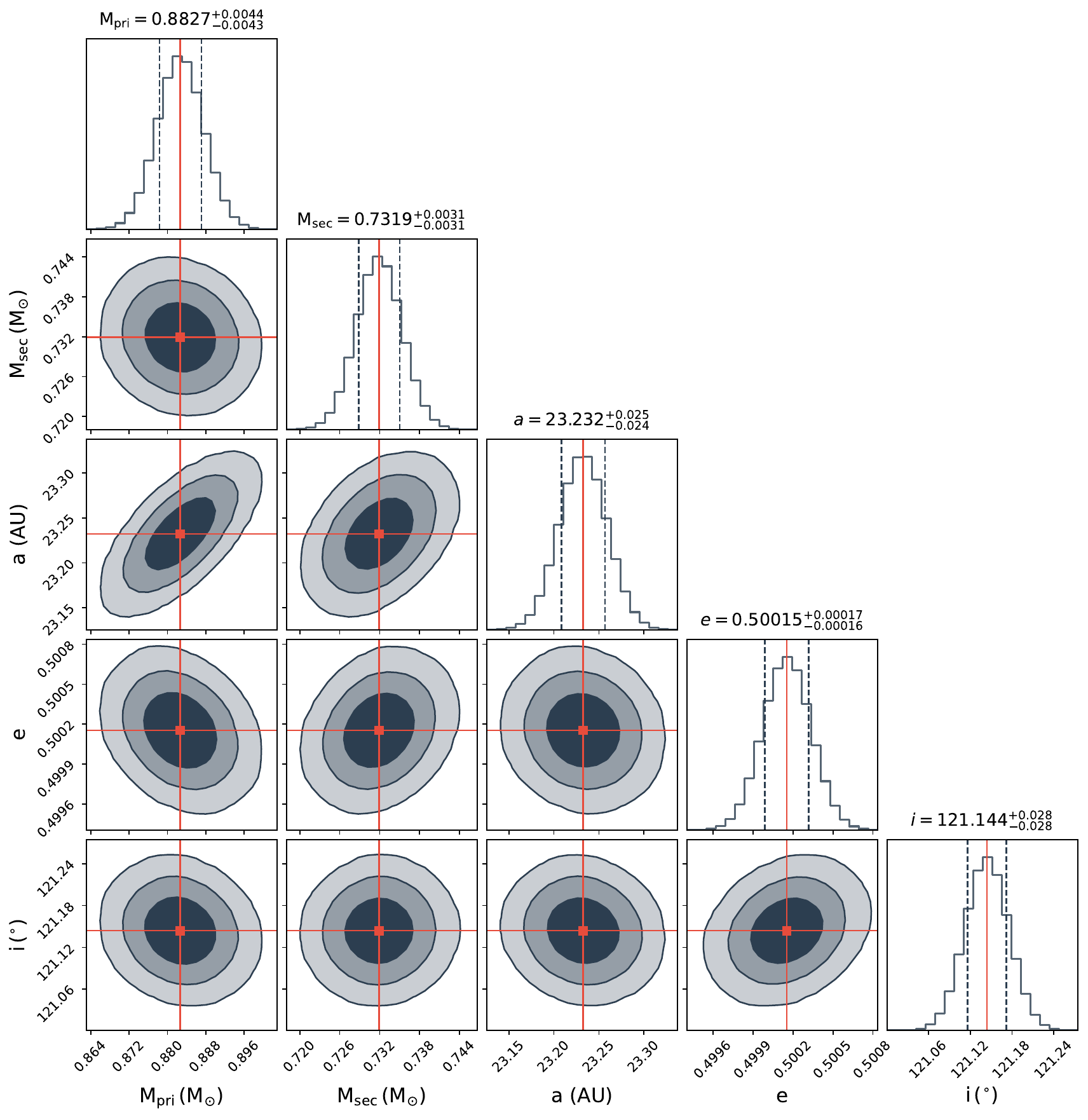}
    \caption{
Posterior distributions of key orbital parameters for the 70~Oph~AB binary from the joint \texttt{orvara} fit.  
The panels show marginalized one- and two-dimensional posteriors for the primary and secondary masses ($M_{\rm pri}$, $M_{\rm sec}$), semimajor axis ($a$), eccentricity ($e$), and inclination ($i$).  
Contours represent 1, 2, and 3$\sigma$ credible regions, while vertical dashed lines mark the median and 68\% confidence intervals.}
    \label{fig:70OphA_corner}
\end{figure*}

\begin{deluxetable*}{lccc}
\tablecaption{MCMC Orbit Fitting Results for 70 Ophiuchi B around 70 Ophiuchi A \label{tab:orbit_results}}
\tablehead{
\colhead{Parameter} & \colhead{Median $\pm$ 68\% CI} & \colhead{95\% CI} & \colhead{Prior}
}
\startdata
\multicolumn{4}{c}{\textit{Fitted Parameters}} \\
\hline
RV jitter (HARPS) (\ms) & $7.86^{+0.13}_{-0.13}$ & (7.605, 8.128) & uniform \\
RV jitter (Keck/HIRES) (\ms) & $7.9^{+4.3}_{-2.3}$ & (4.288, 21.088) & uniform \\
RV jitter (PFS/Magellan) (\ms) & $11.8^{+1.3}_{-1.1}$ & (9.764, 14.719) & uniform \\
RV jitter (New Mills Spectrogram) (\ms) & $27.9^{+14}_{-9.0}$ & (13.129, 61.619) & uniform \\
RV jitter (Mills Spectrogram) (\ms) & $0.72^{+0.73}_{-0.28}$ & (0.307, 3.880) & uniform \\
RV jitter (McDonald Observatory) & $11.11^{+1.0}_{-0.87}$ & (9.461, 13.126) & uniform \\
RV jitter (Victoria Spectrogram) (\ms) & $0.067^{+31}_{-0.067}$ & (0.0, 381.541) & uniform \\
RV jitter (New Victoria Spectrogram) (\ms) & $0.044^{+13}_{-0.044}$ & (0.0, 163.716) & uniform \\
RV jitter (APF/Lick) (\ms) & $13.6^{+1.2}_{-1.1}$ & (11.572, 16.261) & uniform \\
\hline
$M_{\rm pri}$ (\msun) & $0.8827^{+0.0044}_{-0.0043}$ & (0.874, 0.891) & uniform \\
$M_{\rm sec}$ (\msun) & $0.7319^{+0.0031}_{-0.0031}$ & (0.726, 0.738) & uniform \\
Semimajor axis $a$ (au) & $23.232^{+0.025}_{-0.024}$ & (23.185, 23.280) & uniform \\
Semimajor axis $a$ ($\arcsec$) & $4.54332^{+0.00089}_{-0.00088}$ & (4.54157, 4.54508) & -- \\
$\sqrt{e}\,\sin\omega$ & $-0.16094^{+0.00070}_{-0.00070}$ & (-0.162, -0.160) & uniform \\
$\sqrt{e}\,\cos\omega$ & $-0.68866^{+0.00017}_{-0.00017}$ & (-0.689, -0.688) & uniform \\
Inclination $i$ ($^\circ$) & $121.144^{+0.028}_{-0.028}$ & (121.089, 121.199) & $\sin i$ \\
PA of ascending node $\Omega$ ($^\circ$) & $121.666^{+0.032}_{-0.032}$ & (121.604, 121.729) & uniform \\
Mean longitude ($^\circ$) & $299.191^{+0.048}_{-0.048}$ & (299.097, 299.285) & uniform \\
Parallax (mas) & $195.56^{+0.19}_{-0.19}$ & (195.191, 195.929) & Gaussian (Gaia EDR3) \\
\hline
\multicolumn{4}{c}{\textit{Derived Parameters}} \\
\hline
Period (yrs) & $88.126^{+0.010}_{-0.010}$ & (88.106, 88.147) & -- \\
Argument of periastron $\omega$ ($^\circ$) & $193.154^{+0.058}_{-0.058}$ & (193.041, 193.267) & -- \\
Eccentricity $e$ & $0.50015^{+0.00017}_{-0.00016}$ & (0.500, 0.500) & -- \\
Semimajor axis (mas) & $4543.32^{+0.89}_{-0.88}$ & (4541.57, 4545.076) & -- \\
Periastron time $T_0$ (JD) & $2477904.6^{+8.4}_{-8.4}$ & (2477888.157, 2477921.275) & -- \\
Mass ratio $q$ & $0.8292^{+0.0058}_{-0.0057}$ & (0.818, 0.841) & -- \\
\enddata
\end{deluxetable*}

\begin{figure*}
    \centering    \includegraphics[width=1\linewidth]{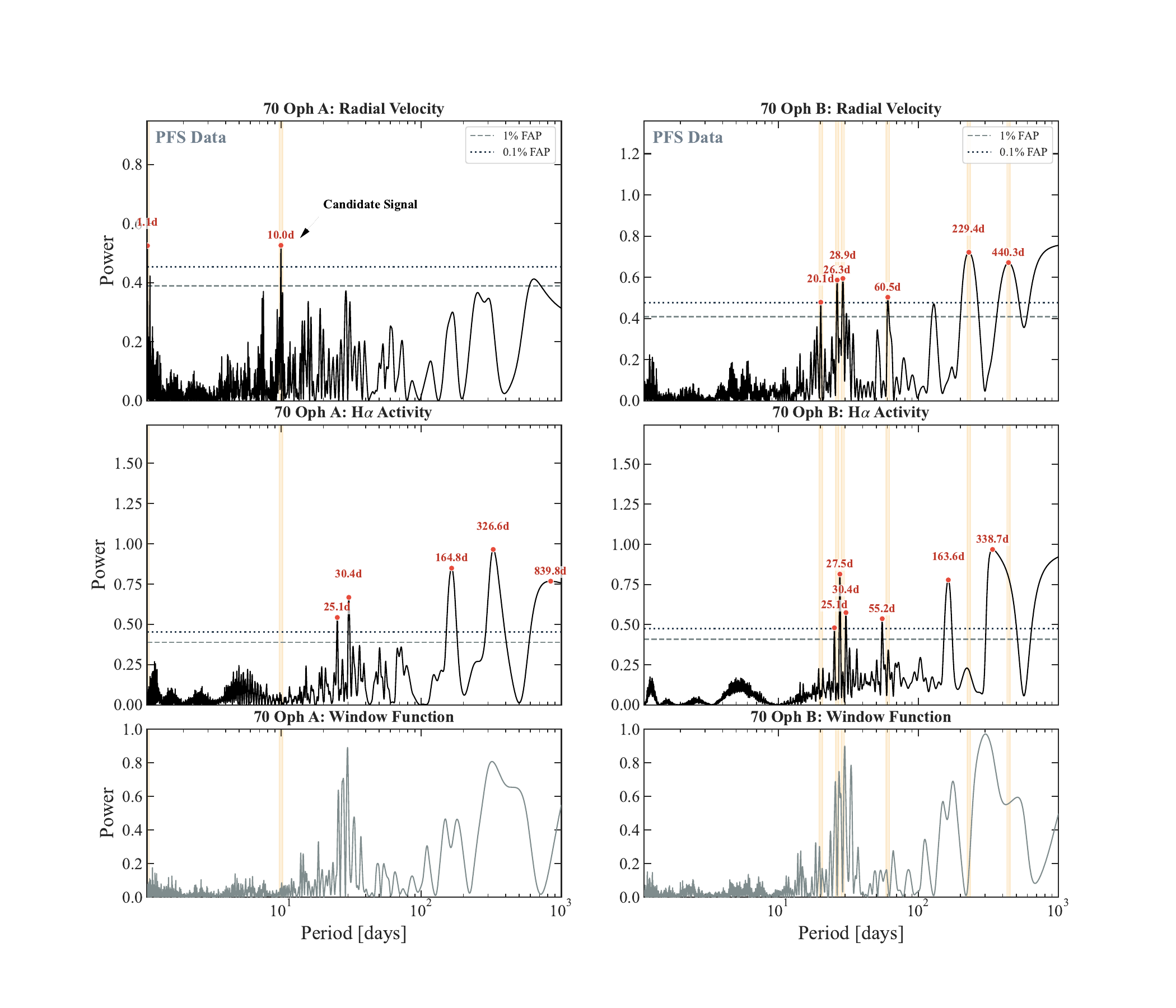}
\caption{Generalized Lomb-Scargle periodograms of the nightly binned PFS radial velocities (top panels), $H_{\alpha}$ indices (middle panels), and spectral window functions (bottom panels) for 70 Oph A (left) and 70 Oph B (right). Note that the long-period binary orbital trend is not subtracted here. The dark-gray dotted and light-gray dashed horizontal lines mark the $0.1\%$ and $1\%$ false-alarm probability (FAP) levels, respectively. Vertical orange shaded bands indicate significant peaks in the RV periodograms above the $0.1\%$ FAP.}
\label{fig:periodogram}
\end{figure*}

\section{Binary Orbit Modeling}\label{sec:orbit}

Thanks to the early discovery of 70~Oph~AB, it has been monitored visually and spectroscopically for over a century. The RV data in total constitute one of the longest radial-velocity baselines available for a nearby binary system. The earliest RV measurements were obtained with the spectrograph at the Dominion Astrophysical Observatory in Victoria, while more recent data were acquired with modern high-precision Doppler instruments, including the Tull coud\'e spectrograph on the 2.7\,m Harlan J.\ Smith Telescope at McDonald Observatory. In this Section, we combine all the available archival astrometric and radial-velocity data with new measurements from PFS observations to refine the binary orbit via a joint Bayesian orbital analysis.

\subsection{Archival RV Data}

We incorporate RV measurements of 70~Oph~A from a total of eight instruments spanning more than a century of observations, from early photographic spectroscopy to modern sub-meter-per-second echelle spectrographs. 
%The full RV time series is provided in Appendix~\ref{app:rvdata}.

The earliest radial velocities were obtained with the original and reconstructed (“new”) Mills spectrographs at Lick Observatory between 1897 and 1932 \citep{Berman_1932}, and were later remeasured by \citet{Batten_1991}. These comprise 31 photographic velocities of 70~Oph~A with typical uncertainties of $\sim$0.5–1\,km\,s$^{-1}$. Subsequent radial velocities were obtained with the Victoria spectrograph at the Dominion Astrophysical Observatory (DAO) between 1966 and 1990 \citep{Batten_1984,Batten_1991}, consisting of 42 measurements from 1966–1983 and an additional 15 measurements through 1990, with similar precision. Together, these archival photographic datasets span 1897–1990. Although lower precision, they are still useful datasets to anchor the long-term phase evolution of the 88 yr binary RV orbit.

The CCD era introduced significant improvements in RV precision. We incorporate 102 RVs of 70~Oph~A obtained with the Tull coud\'e spectrograph on the 2.7\,m Harlan J.\ Smith Telescope at McDonald Observatory as part of the Phase~II and Phase~III planet-search programs. The data were collected over a 27~yr baseline with typical uncertainties of 3--6\,m\,s$^{-1}$ \citep{Wittenmyer2006}.

We further incorporate high-precision HARPS RV observations obtained over six consecutive nights in July 2004 \citep{Carrier2006}. The original campaign comprised 1,758 individual exposures achieving $\sim$1\,m\,s$^{-1}$ precision. For orbital modeling, we adopt nightly averaged velocities corresponding to six independent epochs. Despite the short time baseline, these data provide an important constraint on the velocity semi-amplitude of the B component. We also include five epochs of Keck/HIRES  \citep{Vogt1994} spectra taken with the iodine cell and retrieved from the Keck Observatory Archive, spanning 2014–2018 and achieving median internal uncertainties of $\sim$1\,m\,s$^{-1}$. 

More recent monitoring of 70~Oph~A was carried out with the Levy spectrograph on the Automated Planet Finder (APF) at Lick Observatory \citep{Vogt2014}. We include a total of 76 APF data points obtained between 2014 and 2019 and published in Table~3 of \citet{Hirsch_2021}, which have typical uncertainties of 2--3\,m\,s$^{-1}$. Finally, we incorporate new high-cadence radial velocities we obtained with the Planet Finder Spectrograph (PFS) on the 6.5\,m Magellan Clay telescope. Our observing campaign covered 48 nights between 2023 August and 2025 March (a total baseline of 600 days), yielding 48 nightly-binned velocities for 70~Oph~A and 44 for 70~Oph~B. Individual exposures achieve uncertainties of 0.7–0.8\,m\,s$^{-1}$, while the nightly binned measurements reach a median precision of $\sim$0.27\,m\,s$^{-1}$. Our PFS data represent the highest-precision RV obtained for the 70~Oph system.

When combining multiple RV datasets, instrument-specific velocity zero-points must be accounted for. To do so, we perform a joint fit in \texttt{orvara} \citep{Brandt2021_orvara} in which each instrument is assigned an independent velocity offset $\gamma$ and additive white-noise jitter term $\sigma$ as free parameters. The relative zero-points of the different instruments are then fitted self-consistently as part of the global orbital solution.

\subsection{Archival Astrometry Data}

Besides RV, we incorporate five years of absolute astrometric constraints from the \textit{Hipparcos} mission \citep{ESA1997}, including both the original catalog solution and the revised reduction of \citet{vanLeeuwen2007}. The \textit{Hipparcos} intermediate astrometric data were processed using the \texttt{htof} code \citep{GMBrandt_2021} within \texttt{orvara}. The \textit{Hipparcos}--\textit{Gaia} proper-motion anomaly from the HGCA (DR3 version) \citep{Brandt2021}, in conjunction with the 5-year baseline of \textit{Hipparcos} epoch astrometry, helps resolve the absolute motion of the secondary star’s orbit relative to the primary star. In addition to absolute astrometry, we incorporate 40 epochs of relative visual astrometry spanning more than a century, compiled from the Washington Double Star Catalog (WDS; \citealt{Mason_2001WDS}). The majority of the twentieth-century measurements have large uncertainties of $\sim$0.01$-$0.03\arcsec\ in separation and $\sim$0.1$-$0.2$^\circ$ in position angle. 

Since both components of the binary are resolved by \textit{Gaia}, we also add a relative astrometric measurement from \textit{Gaia} DR3 at epoch 2016 to our astrometric fit. This single-epoch \textit{Gaia} measurement is an order-of-magnitude more precise than historical visual astrometry. Although the \textit{Gaia} point dominates the precision at its epoch, the long temporal baseline provided by historical measurements ought not to be ignored because they trace the orbital period and geometry over a full orbital cycle. 

\subsection{Binary Orbit Fit}

We modeled the binary orbit of 70~Oph~AB using the \texttt{orvara} orbit code, which jointly fits relative astrometry, absolute astrometry, and multi-instrument radial velocities. As noted above, each RV dataset was assigned an independent zero-point offset, and an instrument-specific RV jitter term was added in quadrature to the formal uncertainties. We adopted Gaussian priors on the \textit{Gaia}~DR3 parallax and on the primary mass ($0.89\pm0.02~M_\odot$), with uniform priors on angular orbital parameters. Parameter inference was performed using the parallel-tempering MCMC sampler implemented in \texttt{orvara}, based on \texttt{emcee} \citep{Foreman-Mackey_2013} and its parallel-tempering extension \citep{Vousden_2021}.

Our joint astrometric and radial-velocity fit yields a tightly constrained orbital solution for the 70~Oph~AB binary. The full set of MCMC posteriors for the orbital solution is shown in Table~\ref{tab:orbit_results}. We measure an orbital period of $P = 88.126 \pm 0.010$~yr, eccentricity $e = 0.50015 \pm 0.00017$, and inclination $i = 121.144^\circ \pm 0.028^\circ$. We obtain dynamical masses of $M_{\rm A} = 0.8827 \pm 0.0043\,M_\odot$ and $M_{\rm B} = 0.7319 \pm 0.0031\,M_\odot$, yielding a mass ratio of $q = 0.8292 \pm 0.0058$. The fractional mass uncertainties of 0.49\% and 0.42\% place 70~Oph~AB among the most precisely characterized nearby solar-type binaries.

The astrometric and RV orbits from our joint fit are shown in Figure~\ref{fig:binary_orbit_2}. The visual astrometry is fitted well and tightly anchored by the \textit{Gaia} DR3 measurement near epoch 2016. At the current epoch of 2026, the binary is near maximum separation, with a projected separation of $\rho \sim 6.8\arcsec$ or $\sim$35~AU. The radial-velocity orbit is likewise well constrained: the maximum-likelihood orbit closely follows the RV measurements, and posterior samples are indistinguishable from the best-fit solution.

Figure~\ref{fig:70OphA_corner} shows the posterior distributions, all of which are nearly Gaussian. We see moderate correlations between $M_{\rm A}$, $M_{\rm B}$, and the semimajor axis $a$. This is expected from the Kepler’s third law ($a^3/P^2 \propto M_{\rm A}+M_{\rm B}$). The eccentricity posterior distribution is extremely narrow, centered at $e = 0.50015 \pm 0.00017$, corresponding to a fractional precision of $\sim3\times10^{-4}$. We note that the seemingly near-exact eccentricity value of 0.5 is likely coincidental because it is consistent with previous solutions (e.g., \citealt{Pourbaix2000}) reporting values very close to 0.5.

Our results are remarkably consistent with early orbital determinations reporting $P \simeq 88.3$--88.4~yr and $e \approx 0.50$ \citep{Heintz1988, Batten1991}. We compare our solution to \citet{Eggenberger2008}, who obtained $P = 88.39 \pm 0.03$~yr, $e = 0.495 \pm 0.002$, and masses of $0.89 \pm 0.02\,M_\odot$ and $0.73 \pm 0.01\,M_\odot$ for the primary and secondary, respectively, by combining HARPS RVs with historical visual astrometry and archival RVs from \citet{Pourbaix2000}. Relative to that pre-\textit{Gaia} solution, our period is shorter by 0.264~yr ($8.4\sigma$) and our eccentricity is higher by $2.6\sigma$. The eccentricity uncertainty is reduced by a factor of $\sim$12, from 0.002 to 0.00017. The dynamical masses we found agree with those of \citet{Eggenberger2008} to within $0.6\sigma$ and $1.7\sigma$, but are more precise by factors of $\sim$4.6 and $\sim$3.2. The corresponding signal-to-noise ratios are S/N$\sim$201 and S/N$\sim$236 for the primary and secondary, respectively. In fact, these masses are among the most accurate K dwarfs masses ever measured, surpassed only by $\alpha$~Cen~B (S/N = 363; \citet{Akeson_2021}). These measurements will enable new asteroseismic analyses using updated masses, stellar parameters, abundances, opacities, and evolutionary models, which may help reconcile the age dependency. The mass ratio and eccentricity from our solution fit right in with the known distribution of field binaries at 20--30~AU separations \citep{Raghavan2010,Tokovinin2014,Tokovinin2018}, and they align with typical formation scenarios like turbulent core fragmentation \citep{Offner2010,Bate2012}.

\textcolor{black}{In addition to fitting 70~Oph~B's orbit around the A component, we also do
the reverse for 70~Oph~A's orbit around the B component. The resulting
orbital solution will be used to compute and subtract the RV acceleration
induced by 70~Oph~A from the 70~Oph~B time series in
Section~\ref{sec:companionsearch}. For this fit, we utilize the same
relative astrometry together with the archival radial velocities of
70~Oph~B from the Mills spectrograph at Lick Observatory and the Victoria
spectrograph at DAO, combined with our new $\sim$2-yr PFS radial velocities
of 70~Oph~B. This RV baseline is far too short to independently constrain
the 88-yr binary period, eccentricity, or inclination, so we adopt Gaussian
priors on the dynamical masses $M_A$ and $M_B$ from the posterior we
obtained for A, while allowing all other orbital elements to float. The resulting masses, $M_A = 0.8831 \pm 0.0038\,M_\odot$ and
$M_B = 0.7315 \pm 0.0031\,M_\odot$, are in good agreement with those from
the 70~Oph~A fit ($M_A = 0.8827^{+0.0044}_{-0.0043}\,M_\odot$, $M_B = 0.7319 \pm 0.0031\,M_\odot$; Table~\ref{tab:orbit_results}),
confirming the consistency of the two independent orbital solutions. The full posterior result for this fit is provided
in Appendix~\ref{app:orbit_B}.}

\textcolor{black}{We note that 70~Oph~A and B have \textit{Gaia} DR3 RUWE values of 3.27 
and 3.71, respectively, which are roughly $2.3$--$2.6$ times above the canonical threshold of 1.4 \citep{GaiaDR3}. This is expected because at $V \sim 4$, both components lie near \textit{Gaia}'s saturation limit, where 
astrometric residuals are systematically inflated, and the single-star 
pipeline does not account for photocentre displacement from the companion. 
Nonetheless, our solution is anchored primarily by long-baseline RV data 
and historical visual astrometry rather than \textit{Gaia} alone, and its 
consistency with pre-\textit{Gaia} results \citep{Eggenberger2008} 
confirms that these systematics do not bias our orbital parameters.}

\begin{figure*}
\centering
\includegraphics[width=\linewidth]{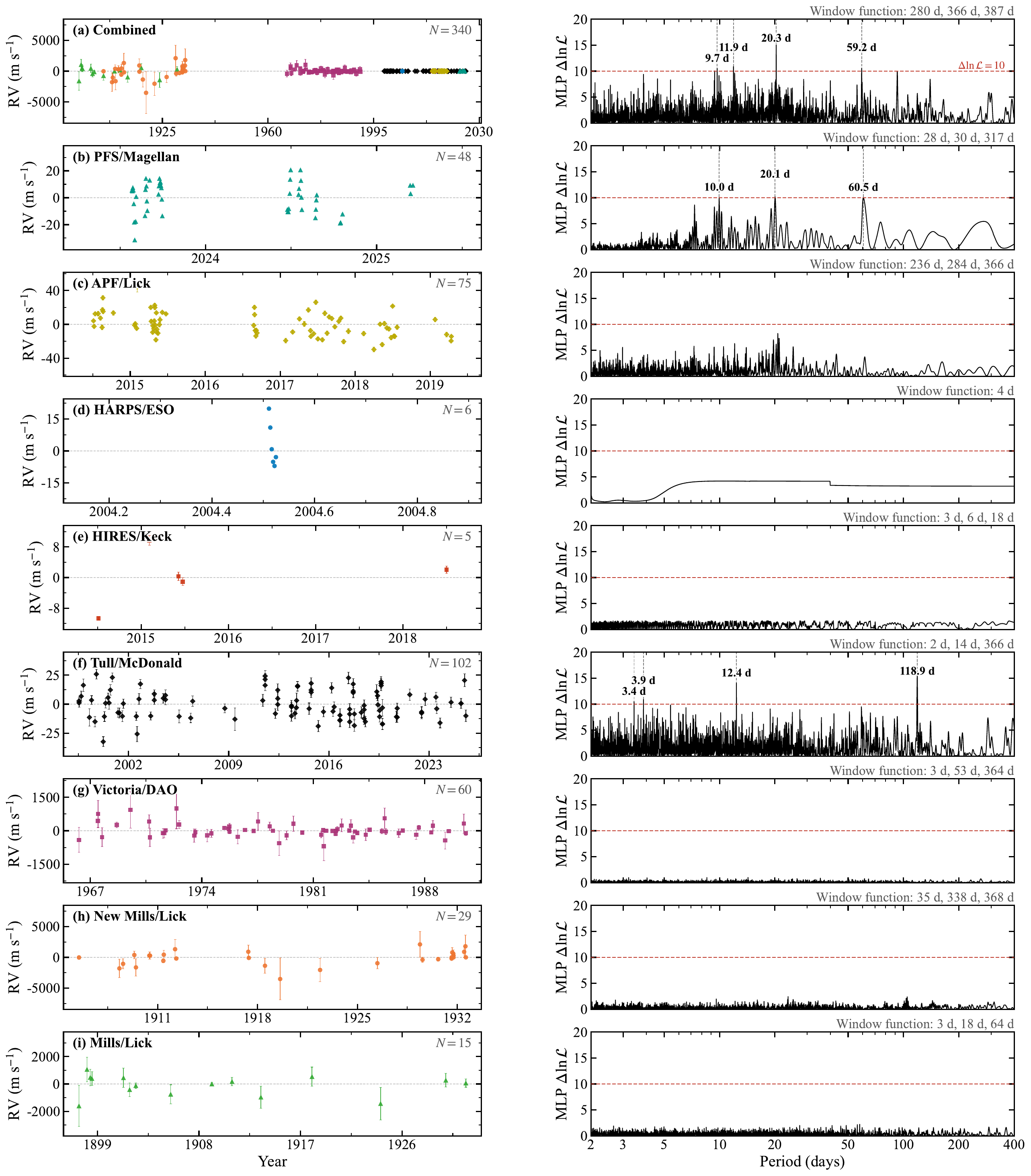}
\caption{\textcolor{black}{Maximum likelihood periodograms (MLP) of the combined and individual 70~Oph~A RV 
residuals after removing the acceleration of 70~Oph~B. Per-instrument offsets and jitter 
are simultaneously fit at each trial frequency. The red dashed line marks the 1\% FAP 
threshold ($\Delta\ln\mathcal{L} = 10$). Dominant spectral window peaks are noted above 
each panel. Gray dotted lines mark signals of interest at $9.7$, $11.9$, $20.3$, and 
$59.2$~days.}}
\label{fig:mlp_all}
\end{figure*}

\section{Search for Planetary Companions}
\label{sec:companionsearch}

\subsection{Periodograms}
\label{sec:periodograms}

\subsubsection{GLS Periodograms}

Given the wealth of available RV data, we conduct a search for planets in the 70~Oph system using GLS periodograms. The PFS dataset shown in Figure~\ref{fig:nightly_RVs} provides the highest RV precision ($\sim$0.7--0.8~m~s$^{-1}$) in all RV datasets, followed by APF ($\sim$1--2~m~s$^{-1}$) and McDonald Observatory data ($\sim$2--3~m~s$^{-1}$), while older archival data have substantially larger uncertainties. We therefore begin by analyzing a GLS periodogram of the nightly binned PFS data, which is well suited for analyzing unevenly sampled time-series observations.

Figure~\ref{fig:periodogram} shows the GLS periodograms of the nightly binned PFS RV dataset and the H$\alpha$ activity indicator for both components. For 70~Oph~A, the only RV peak exceeding the 0.1\% false-alarm probability (FAP) occurs at a period of $\sim$10.0~days. Although this candidate signal is absent in the H$\alpha$ periodogram, its coincidence with the first harmonic ($P_{\text{rot}}/2$) of the $\sim$20-day stellar rotation period is immediately apparent \citep{Noyes1984, Ayres2022}. The fundamental rotation period at $\sim$20~days is also detected. No other RV signals exceed the 0.1\% FAP threshold.

The GLS periodogram of 70~Oph~B exhibits a more complex and noisier structure than that of the primary, indicating higher activity levels. A cluster of peaks between 20 and 30~days, along with excess power in H$\alpha$ over the same range reflects rotational modulation from multiple active regions and differential rotation. Notably, both components show a sharp $\sim$20-day RV peak that is absent in H$\alpha$. Given that a Keplerian signal would not be expected to appear at the same period in both stars of this wide binary, the shared $\sim$20-day peak further supports an activity-related interpretation for the 20-day signal. \textcolor{black}{Additional power in the 70~Oph~B periodogram near 60~days and between 200-450~days is present in this GLS. However, we show in Section~\ref{sec:companionsearch} that this power does not persist once the binary-induced radial velocity trend is removed, indicating that it arises from the interaction of the long-period binary trend with the observing window rather than from an independent periodic signal.}

%One possible explanation is that RVs trace localized, long-lived photospheric spots, whereas H$\alpha$ traces more diffuse, rapidly evolving chromospheric plages. The shared $\sim$20-day period suggests that both stars may rotate with the same period, as studies have shown that rotation periods in a binary system tend to converge as the stars age and their spin-down rates slow \citep{vanSaders_2016}. 

\subsubsection{MLP Periodograms}

%Since the only candidate signal appears at $\sim$10~days in 70~Oph~A, we hereafter focus our planet search on the A component. 

So far, the periodogram analysis presented above did not account for the long-period
binary orbital trend. We now subtract this trend from the RV time series of
each component using the best-fit orbital solution shown in
Figure~\ref{fig:binary_orbit}: for 70~Oph~A, we remove the RV contribution
induced by 70~Oph~B, and for 70~Oph~B, we remove the RV contribution induced
by 70~Oph~A. We then recompute the periodograms of the binary-subtracted
residuals for both components. For this analysis, we adopt maximum likelihood periodograms
\citep[MLP;][]{Baluev_2008,Zechmeister2019} instead of the GLS, as the GLS assumes a common RV offset across all datasets and does not account for instrument-specific jitter. The MLP simultaneously fits individual offsets
and jitter terms at each trial frequency, making it more appropriate for our heterogeneous multi-instrument dataset.
%The resulting residuals (O--C) for each instrument are already on a common reference frame because the \texttt{orvara} orbital fit included per-instrument velocity zero points. Therefore, no further RV offset corrections are required to combine the datasets.

Figure~\ref{fig:mlp_all} presents the MLP periodograms for 70~Oph~A after 
removing the orbital contribution of 70~Oph~B. The combined RV dataset is dominated 
by the high-precision PFS and McDonald observations, and its periodogram reveals four 
prominent signals at 9.7, 11.9, 20.3, and 59.2~days. The $\sim$20.3-day signal is 
firmly detected as the stellar rotation period ($P_{\rm rot}$), appearing as the 
strongest peak in both the McDonald and PFS datasets. We discuss the significance of the 
remaining signals below.

%Figure~\ref{fig:instrument_gls} presents the GLS periodograms for 70~Oph~A after removing the orbital contribution of 70~Oph~B. The combined RV dataset is dominated by the high-precision PFS and McDonald observations, and its periodogram reveals five prominent signals at 7.3, 9.4, 12.1, 20.3, and 59.2~days. The $\sim$20.3-day signal is firmly detected as the stellar rotation period ($P_{\rm rot}$), appearing as the strongest peak in both the McDonald and PFS datasets. We discuss the significance of the remaining signals below.

The $\sim$9.7-day signal ($P_{\rm rot}/2$) is our primary candidate 
of interest. This signal is present in all the high-precision datasets and is 
particularly prominent in the 2-year PFS data relative to the 20-year McDonald dataset, 
despite the temporal overlap between the two. We propose two possible explanations. 
If the signal originates from an orbiting planet, the higher precision of PFS may enable 
such a low-amplitude signal of a few m~s$^{-1}$ to be detected more readily than in the 
lower-precision McDonald data. Alternatively, if the signal is instead the first harmonic 
of the rotation period, variations in amplitude and phase coherence over decadal 
timescales could manifest as reduced power in the longer McDonald dataset 
\citep{Haywood_2014}.

%such a low-amplitude signal to be detected more clearly than in the lower-precision McDonald data. Alternatively, if the signal is instead the first harmonic of the rotation period, variations in amplitude and phase coherence over decadal timescales could reduce its power in the longer McDonald dataset \citep{Haywood_2014}.

\begin{figure*}[htbp]
  \centering
  
  % --- First Image ---
  \begin{minipage}[t]{0.48\linewidth}
    \centering
    \includegraphics[width=0.93\linewidth]{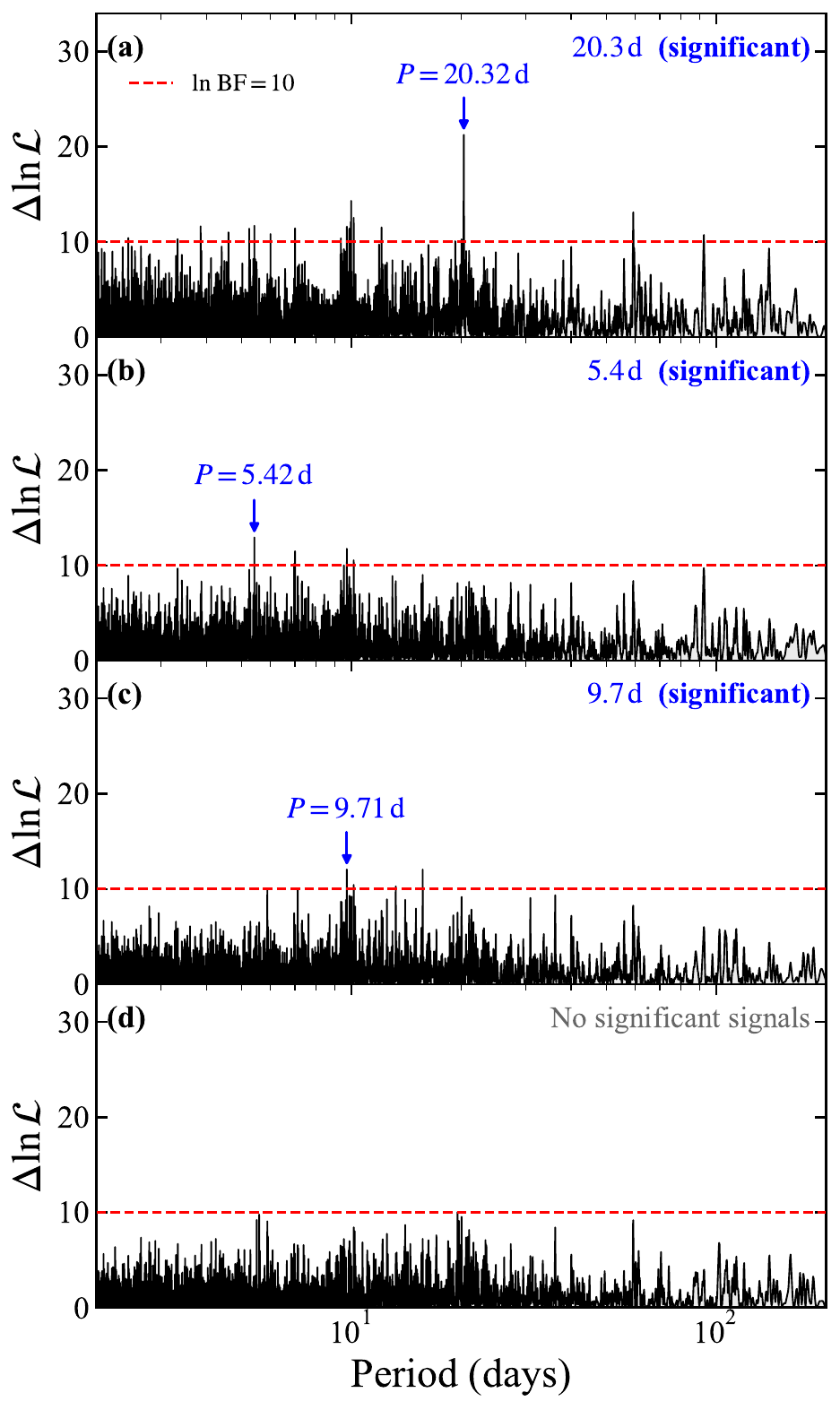}
       \caption{Maximum-likelihood periodograms (MLPs) of 70~Oph~A after subtracting the binary acc. Significant periodicities are recovered at 20.3~days ($P_{\rm rot}$), 9.7~days ($P_{\rm rot}/2$), and 5.4~days ($P_{\rm rot}/3$). After sequential removal of these signals, no additional significant periodicities appear in the data.}
    \label{fig:mlp_sequential}
  \end{minipage}
  \hfill % This pushes the two minipages to the edges
  % --- Second Image ---
  \begin{minipage}[t]{0.48\linewidth}
    \centering
    \includegraphics[width=0.956\linewidth]{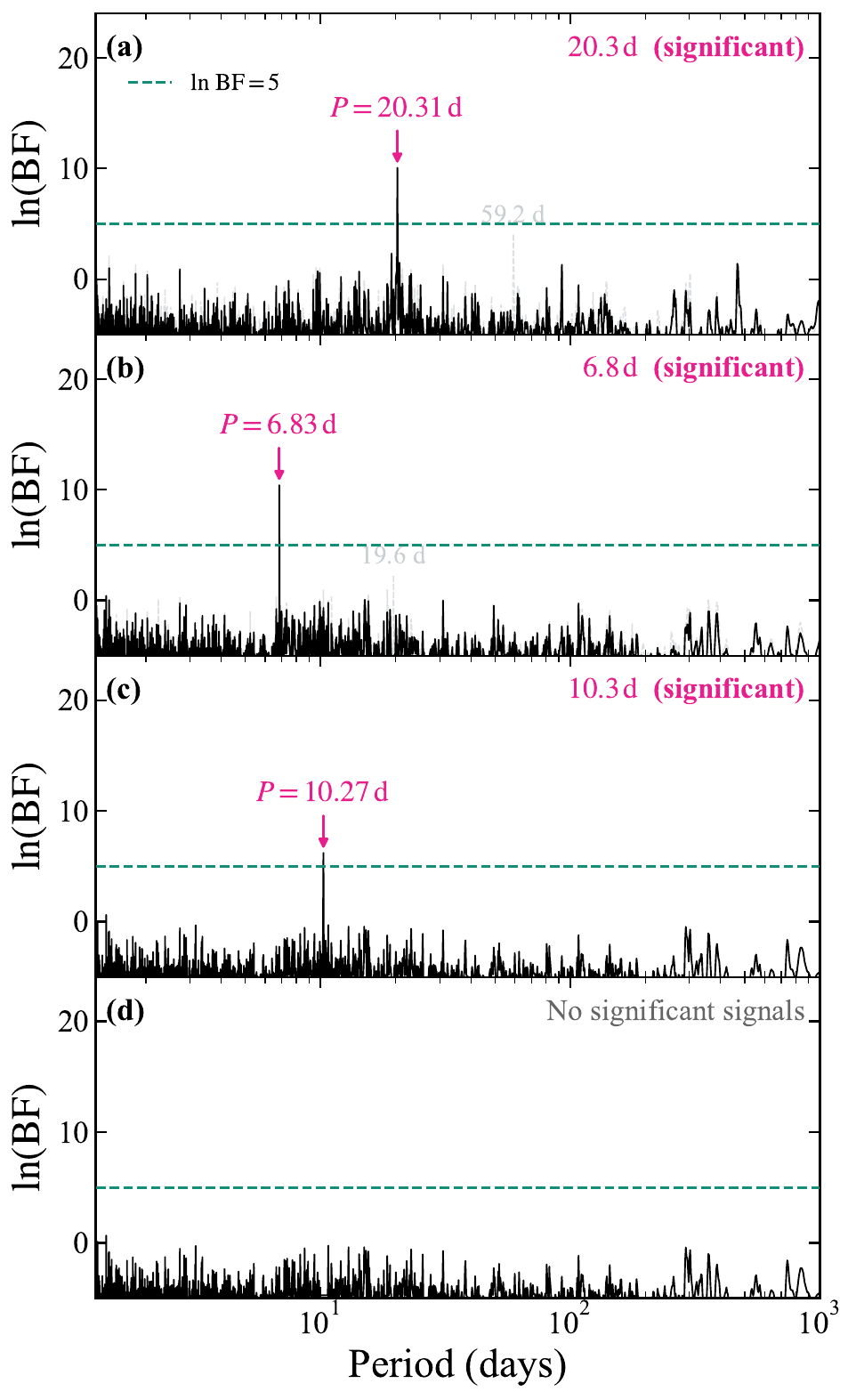}
    \caption{Bayes factor periodograms (BFPs) of binary-subtracted RV residuals of 70~Oph~A from \texttt{Agatha}. Each panel shows the BFP after removing the strongest signal identified in the previous iteration. Three significant signals are found at 20.3 days ($P_{\rm rot}$), 10.2 days ($P_{\rm rot}/2$), and 6.9 days ($P_{\rm rot}/3$).}
    \label{fig:agatha}
  \end{minipage}
  
\end{figure*}

%Next, a 7.3-day signal is detected in the PFS, McDonald, and APF periodograms. However, this period is close to $\sim P_{\rm rot}/3$ and corresponds to the third rotational harmonic. 

\begin{figure}[htbp]
\centering
\includegraphics[width=\linewidth]{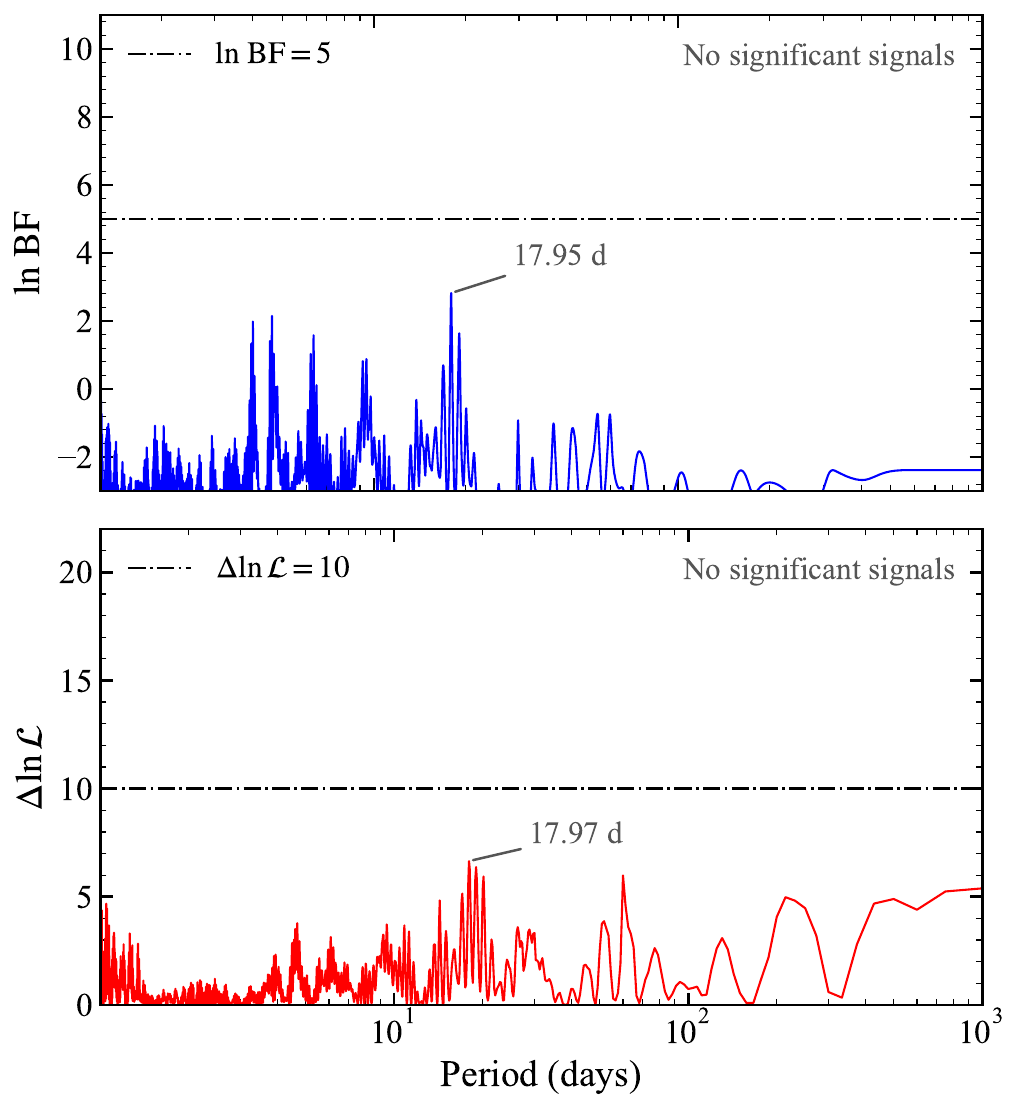}
\caption{Periodogram analysis of 70~Oph~B PFS radial velocity residuals after subtracting the binary acceleration. Top: \texttt{Agatha} Bayes factor periodogram where the dash-dotted line marks $\ln\,\mathrm{BF} = 5$.
Bottom: Maximum Likelihood Periodogram (MLP) where the dash-dotted line marks $\Delta\ln\mathcal{L} = 10$. Neither method detects a significant periodic signal in the RV data of 70~Oph~B.} \label{fig:periodogram_B}
  \end{figure}

We also find additional excess power at 11.9 and 59.2~days. The 59.2-day signal is detected only in the PFS data. Furthermore, the 59.2-day peak appeared in the PFS periodograms of both 70~Oph~A and B in our previous analysis (Figure~\ref{fig:periodogram}) which sits at $3\times P_{\rm rot}$. Therefore, we favor its interpretation as a rotational alias or harmonic of the $\sim$20-day signal. Finally, the 11.9-day signal is strong in the McDonald data ($\ge$ 1\% FAP) but is only somewhat present in PFS ($\sim$1\% FAP). This period does not correspond to a simple integer multiple or fraction of $P_{\rm rot}$, and its origin remains unclear. It may reflect additional structure in the rotational variability. To test its stability, we split the data by half and found that the signal is not consistently recovered. The lack of coherence favors a complex rotation signal over a Keplerian signal due to a planet.

Next, we iteratively fit and subtract sinusoidal models for the dominant signals from the 
combined RV dataset for 70~Oph~A using the MLP until no significant power remains above the threshold of $\Delta\ln\mathcal{L} = 10$. As shown in Figure~\ref{fig:mlp_sequential}, the strongest peak in the combined residuals occurs at 20.3~days, consistent with the stellar rotation period ($P_{\rm rot} \approx 20$~days). After removing this component, the next 
strongest signal appears at 5.4~days, followed by 9.7~days ($P_{\rm rot}/2$). After subtracting all three terms, no significant power remains above $\Delta\ln\mathcal{L} = 10$. The emergence of the 5.4-day signal ($P_{\rm rot}/4$) and 9.7-day signal ($P_{\rm rot}/2$) after removal of the rotation period is consistent with higher-order rotational harmonics.

We repeat the same MLP analysis for 70~Oph~B by removing the RV trend
induced by the A component, using the orbital solution derived in
Appendix~\ref{app:orbit_B}. As shown in Figure~\ref{fig:periodogram_B}, the
MLP does not detect a significant periodic signal, reaching a peak
$\Delta\ln\mathcal{L}$ of only $\sim$6 near 18~days, below the
significance threshold of 10. This marginal $\sim$18-day feature is roughly at half of the $\sim$34-day rotation period reported
for 70~Oph~B by \citet{Baliunas1996} from Mount Wilson $S$-index
monitoring (see Section~\ref{sec:stellarproperty}), and we speculate that
it may represent the first rotational harmonic for 70~Oph~B. We thus conclude our MLP periodogram analysis at this stage.

%As shown in Figure~\ref{fig:gls_all},the strongest residual peaks after removal of the binary trend occur at 9.3 and 20.3~days, consistent with the stellar rotation period ($P_{\text{rot}} \approx 20$~days) and its first harmonic. Removing these components suppresses the power at 12.1 and 59.2~days, indicating that both signals are likely aliases. The next strongest peak at 7.3~days, approximately $P_{\text{rot}}/3$, is consistent with a higher-order rotational harmonic. After subtracting this term, no significant power remains above the noise floor, aside from residual structure introduced by imperfect sinusoidal modeling. We therefore conclude the GLS analysis at this stage.

\subsubsection{Agatha Periodograms}
To better characterize the potential signals in the periodogram, we further employ the \texttt{Agatha} Bayes factor periodograms (BFPs) described in \citet{Feng_2017}. The primary advantage of \texttt{Agatha} is that it evaluates the statistical significance of candidate periodic signals by modeling time-correlated noise through a moving-average (MA) framework, rather than simply identifying peaks in power. Unlike simple sinusoidal fits, \texttt{Agatha} assesses whether the inclusion of an additional periodic signal is favored by the data using the Bayesian Information Criterion (BIC). Although not a replacement for a full MCMC Bayesian analysis, it provides a first-order, evidence-based comparison of models with different numbers of periodic components and evaluates each signal in the presence of others rather than through sequential subtraction.

Figure~\ref{fig:agatha} shows the recursive Bayes factor periodograms for 70~Oph~A, computed using a first-order moving-average model (MA=1) and an oversampling factor of 15. We identify three significant signals at 20.3, 10.3, and 6.8~days, all exceeding a Bayes factor threshold of $\ln \rm BF > 5$. These results are consistent with our MLP findings (20.3, 9.7, and 5.4~days) and correspond closely to $1\times P_{\rm rot}$, $1/2\times P_{\rm rot}$, and $1/3\times P_{\rm rot}$. We note that some residual power remains at 19.6~days, which we interpret as evidence of evolving active regions or differential rotation. No further independent signals exceed the adopted Bayes factor threshold. We also recover a 59.2-day signal with a Bayes factor marginally below 5. We interpret this periodicity as a rotational alias near $3\times P_{\text{rot}}$, and therefore do not repeat the detailed discussion here.

We similarly compute the Agatha BFP for the binary-subtracted 70~Oph~B
residuals. As shown in Figure~\ref{fig:periodogram_B}, no signal exceeds
the adopted Bayes factor threshold of $\ln\text{BF} = 5$. The highest peak
reaches only $\ln\text{BF}\sim3$, near 17.95~days, consistent with the
marginal $\sim$18-day feature identified in the MLP analysis
(Section~\ref{sec:companionsearch}). We therefore find no significant
periodic signal in the 70~Oph~B radial velocities using either method.

%The slight discrepancies in the periods of the shorter harmonics likely arise from the differing treatments of stellar activity noise within the Bayesian framework of \texttt{Agatha} compared to the frequentist GLS approach. 

\subsection{Summary of Periodogram Analyses}

Table~\ref{tab:signals} lists the periodic signals identified throughout our periodogram analysis. For 70~Oph~B, neither the MLP nor the Agatha BFP detects a significant signal. The only marginal feature at $\sim$18~days may correspond to the first harmonic of the star's rotation period (Section~\ref{sec:companionsearch}). For 70~Oph~A, we consistently identify three signals across methods and datasets: the stellar rotation period and its first two harmonics ($P_{\text{rot}} \approx 20$~d, $P_{\text{rot}}/2 \approx 10$~d, and $P_{\text{rot}}/3 \approx 7$~d), with the 10-day signal being the most prominent. The 11.9-day and 59.2-day signals are likely transient features or sampling aliases. In Section~\ref{sec:GPR}, we apply Gaussian Process Regression to disentangle potential planetary signals from stellar activity in the RV data of 70~Oph~A, testing the 7, 10, 12, and 60-day signals from Table~\ref{tab:signals}, since these are the periods whose origin as rotational harmonics versus genuine Keplerian signals remains ambiguous from the periodogram analysis alone.

We do not include visual astrometry in the fit, as the typical uncertainties of the available measurements (of order tens of milliarcseconds) are far larger than the expected astrometric reflex motion induced by any putative inner companion. For example, a sub-Jupiter-mass companion on a $\sim$10-day orbit around 70~Oph~A would induce astrometric signatures at the $\lesssim 0.1$~mas level, well below the precision of the available visual astrometric data and challenging even for current space-based astrometry.

\section{Gaussian-Process Regression Modeling}
\label{sec:GPR}

Since our periodogram analysis reveals no significant signals in 70~Oph~B, 
we hereafter focus our GPR analysis and planet search on the A component. We use the \texttt{RadVel} software package \citep{Fulton_2018} to model the RV data of 70~Oph~A. Each Keplerian signal is described by the standard orbital parameters, and we reparameterize the eccentricity as $\sqrt{e}\sin\omega$ and $\sqrt{e}\cos\omega$ to improve sampling efficiency and mitigate degeneracies near $e=0$ \citep{Rosenthal_2021}. Although the global systemic velocity and long-term acceleration have been 
removed via subtraction of the best-fit binary orbital solution, we still include per-instrument velocity offsets $\gamma$ and jitter terms $\sigma$ 
to account for instrumental zero-point differences and excess white noise 
between instruments, respectively. These parameters are fit simultaneously 
with the Keplerian and activity model parameters. %We do not include systemic velocity offsets or long-term acceleration terms, as these components have already been removed when we subtracted the best-fit binary orbital solution. However, we allow for independent RV jitter terms for each instrument, which are fit simultaneously with the Keplerian and activity model parameters.

We adopt the Gaussian process regression (GPR) method \citep{Aigrain_2023}, which is commonly used to model quasi-periodic stellar variability when activity amplitudes are comparable to planetary signals \citep{Haywood_2014,Nava_2020}. We utilize a quasi-periodic covariance kernel, which captures both rotational modulation and the finite coherence timescale of active regions  \citep[e.g.,][]{Blunt_2023}. % To avoid overfitting, we restrict our GPR analysis to a subset of the data with higher cadence and precision (i.e. APF, McDonald, and PFS).

All of the GP models we fit employ a quasi-periodic covariance kernel of the form
\begin{equation}
k(t_i,t_j)=A^2
\exp\!\left[
 -\frac{(t_i-t_j)^2}{2\tau^2}
 -\frac{\sin^2\!\left(\pi (t_i-t_j)/P_{\rm rot}\right)}{2w^2}
\right],
\end{equation}
where $A$ is the amplitude of the correlated variability, $\tau$ is the exponential decay timescale that governs the evolution of active regions, $P_{\rm rot}$ is the stellar rotation period, and $w$ controls the coherence of the periodic component \citep{Haywood_2014,Nava_2020}. The RV and activity periodograms show strong, narrow power at the stellar rotation period and its harmonics, with no additional long-term structure requiring a complex activity model. Thus, we adopt the quasi-periodic kernel as a simple model that captures the rotation-driven variability.

%More flexible kernels do not improve the fit statistically and instead increase model complexity and the risk of overfitting. 

\begin{table*}[ht]
\caption{Lists of periodic signals in 70~Oph~AB radial velocity periodogram analysis and their interpretations.}
\label{tab:signals}
\begin{tabular}{llccccl}
\hline\hline
Star & Dataset & Method & Period (d) & Significance & FAP & Interpretation \\
\hline
70~Oph~A & Combined   & MLP & 9.7   & $\Delta\ln\mathcal{L} > 10$ & $<1\%$ & $P_{\rm rot}/2$ harmonic \\
         & Combined   & MLP & 11.9  & $\Delta\ln\mathcal{L} > 10$ & $<1\%$ & \textbf{Uncertain} \\
         & Combined   & MLP & 20.3  & $\Delta\ln\mathcal{L} > 10$ & $<1\%$ & Stellar rotation ($P_{\rm rot}$) \\
         & Combined   & MLP & 59.2  & $\Delta\ln\mathcal{L} > 10$ & $<1\%$ & $3\times P_{\rm rot}$ alias \\
         & McDonald   & MLP & 3.4   & $\Delta\ln\mathcal{L} > 10$ & $<1\%$ & Window function alias \\
         & McDonald   & MLP & 3.9   & $\Delta\ln\mathcal{L} > 10$ & $<1\%$ & Window function alias \\
         & McDonald   & MLP & 12.4  & $\Delta\ln\mathcal{L} > 10$ & $<1\%$ & \textbf{Uncertain}  \\
         & McDonald   & MLP & 118.9 & $\Delta\ln\mathcal{L} > 10$ & $<1\%$ & Window function alias \\
         & PFS        & MLP & 10.0  & $\Delta\ln\mathcal{L} > 10$ & $<1\%$ & $P_{\rm rot}/2$ harmonic \\
         & PFS        & MLP & 20.1  & $\Delta\ln\mathcal{L} > 10$ & $<1\%$ & Stellar rotation ($P_{\rm rot}$) \\
         & PFS        & MLP & 60.5  & $\Delta\ln\mathcal{L} > 10$ & $<1\%$ & $3\times P_{\rm rot}$ alias \\
\hline
70~Oph~A & Sequential & MLP & 20.32 & $\Delta\ln\mathcal{L} > 10$ & $<1\%$ & Stellar rotation ($P_{\rm rot}$) \\
(pre-whitening) & Sequential & MLP &  5.42 & $\Delta\ln\mathcal{L} > 10$ & $<1\%$ & $P_{\rm rot}/3$ harmonic \\
         & Sequential & MLP &  9.71 & $\Delta\ln\mathcal{L} > 10$ & $<1\%$ & $P_{\rm rot}/2$ harmonic \\
         & Sequential & BFP & 20.31 & $\ln{\rm BF} > 5$ & $-$ & Stellar rotation ($P_{\rm rot}$) \\
         & Sequential & BFP &  6.83 & $\ln{\rm BF} > 5$ & $-$ & $P_{\rm rot}/3$ harmonic \\
         & Sequential & BFP & 10.27 & $\ln{\rm BF} > 5$ & $-$ & $P_{\rm rot}/2$ harmonic \\
\hline
70~Oph~B & PFS & MLP/BFP & \multicolumn{3}{c}{No significant signals} & $-$ \\
\hline\hline
\end{tabular}
\begin{flushleft}
\footnotesize \textbf{Notes:} We adopt a MLP significance threshold of $\Delta\ln\mathcal{L} = 10$ (FAP $= 1\%$), and a BFP significance threshold of $\ln{\rm BF} = 5$. FAP is not directly defined for the BFP method. The pre-whitening analysis sequentially removes the strongest signal at each iteration.
\end{flushleft}
\end{table*}

\begin{table*}
\centering
\caption{Priors Adopted for Keplerian and Gaussian-Process Models of 70~Oph~A}
\label{tab:priors_gpr}
\begin{tabular}{llll}
\hline\hline
Parameter & Prior & Units & Description \\
\hline
\multicolumn{4}{c}{\textbf{Keplerian Parameters (Candidate Periodic Signals)}} \\
\hline
$P_i$ & $\mathcal{U}(P_{i,\min}, P_{i,\max})$ & days & Orbital period \\
$T_{c,i}$ & $\mathcal{U}(t_0 - P_i/2,\ t_0 + P_i/2)$ & BJD & Time of inferior conjunction \\
$\sqrt{e_i}\cos\omega_i$ & $\mathcal{U}(-1,1)$ & -- & Eccentricity reparameterization$^{a}$ \\
$\sqrt{e_i}\sin\omega_i$ & $\mathcal{U}(-1,1)$ & -- & Eccentricity reparameterization$^{a}$ \\
$K_i$ & $\mathcal{U}(-10.0, 10.0)$ & m s$^{-1}$ & Signed RV semi-amplitude \\
\hline
\multicolumn{4}{c}{\textbf{Gaussian-Process Hyperparameters (Quasi-periodic Kernel)}} \\
\hline
$A$ & $\mathcal{J}(0.01, 100.0)$ & m s$^{-1}$ & GP amplitude \\
$\tau$ & $\mathcal{J}(1.0, 10^4)$ & days & Active-region evolution timescale \\
$P_{\rm rot}$ & $\mathcal{U}(18.0, 25.0)$ & days & Stellar rotation period \\
$w$ & $\mathcal{U}(0.05, 0.6)$ & -- & Periodic coherence scale \\
\hline
\multicolumn{4}{c}{\textbf{Instrumental Parameters}} \\
\hline
$\gamma_{\rm inst}$ & $\mathcal{U}(-100, 100)$ & m s$^{-1}$ & RV zero-point offset (per instrument) \\
$\sigma_{\rm inst}$ & $\mathcal{U}(0.01, 100)$ & m s$^{-1}$ & RV jitter term (per instrument) \\
\hline
\end{tabular}
\begin{flushleft}
\footnotesize
\textbf{Notes:} $\mathcal{U}(a,b)$ denotes a uniform prior between $a$ and $b$, and
$\mathcal{J}(a,b)$ denotes a Jeffreys prior between $a$ and $b$.
Identical Keplerian priors are adopted for each tested periodic signal.
The period prior bounds $(P_{i,\min}, P_{i,\max})$ are informed by the GLS periodogram analysis.
$^{a}$Eccentricity parameters are fixed to zero for the circular Keplerian models.
\end{flushleft}
\end{table*}

We explore the posterior parameter space using a Markov Chain Monte Carlo (MCMC) approach, implemented via \texttt{RadVel}, to jointly sample the Keplerian and stellar-activity model parameters. We adopt broad, weakly informative priors to allow the data to drive the inference \citep{Angus_2018}, specifically utilizing Jeffreys priors for parameters spanning several orders of magnitude, such as the RV semi-amplitude and GP amplitude terms \citep{Figueiredo_2001}. All other parameters are assigned uniform priors. 

For the Keplerian components, orbital periods and times of conjunction are informed by our periodogram analysis. The RV semi-amplitude is restricted between 0.01 and 10~m~s$^{-1}$, as the long temporal baseline ($>20$~yr) and modest RV scatter preclude higher-amplitude signals. In the activity model, the periodic hyperparameter is constrained to a physically motivated range centered on the expected stellar rotation period. We adopt broad priors for the remaining GP hyperparameters governing the amplitude, coherence, and evolutionary timescales. A complete list of these priors is provided in Table~\ref{tab:priors_gpr}.

Even though we subtracted the binary orbit, we still include small velocity offsets and white-noise jitter terms for each instrument. We note that because the signals of interest have periods much shorter than the binary period, subtracting the binary trend has little effect on the analysis. One caveat of our fit is that we allow the RV semi-amplitude to vary symmetrically about zero in the sampling. Negative values are degenerate with a phase shift of $\pi$ and correspond to the same physical solution as positive values. In practice, this choice does not affect our inferred constraints. 

We test five candidate periodic signals in total: three at 20.2, 10.3, and 6.8~days, which are suspected to originate from stellar activity, and two additional signals at 12.1 and 59.2~days. Although we have discussed these signals being rotational harmonics or sampling aliases, we formally test them using Gaussian Process Regression (GPR) to enable a rigorous statistical comparison. 

We restrict our GPR analysis to modern Doppler RV data (post-1980), including PFS (48 measurements spanning 1.6~years), APF (75 measurements spanning 2014–2019 with typical 2–3~m\,s$^{-1}$ precision), McDonald (102 measurements spanning 27~years), and Keck/HIRES (5 measurements), for a total of 230 RV measurements over a 27-year baseline. The best-fit binary orbital solution is subtracted prior to fitting. By comparing models that incorporate quasi-periodic stellar activity with various Keplerian configurations, we statistically assess the most likely origin of each detected periodicity.

\begin{figure*}
\centering
\includegraphics[width=0.9\linewidth]{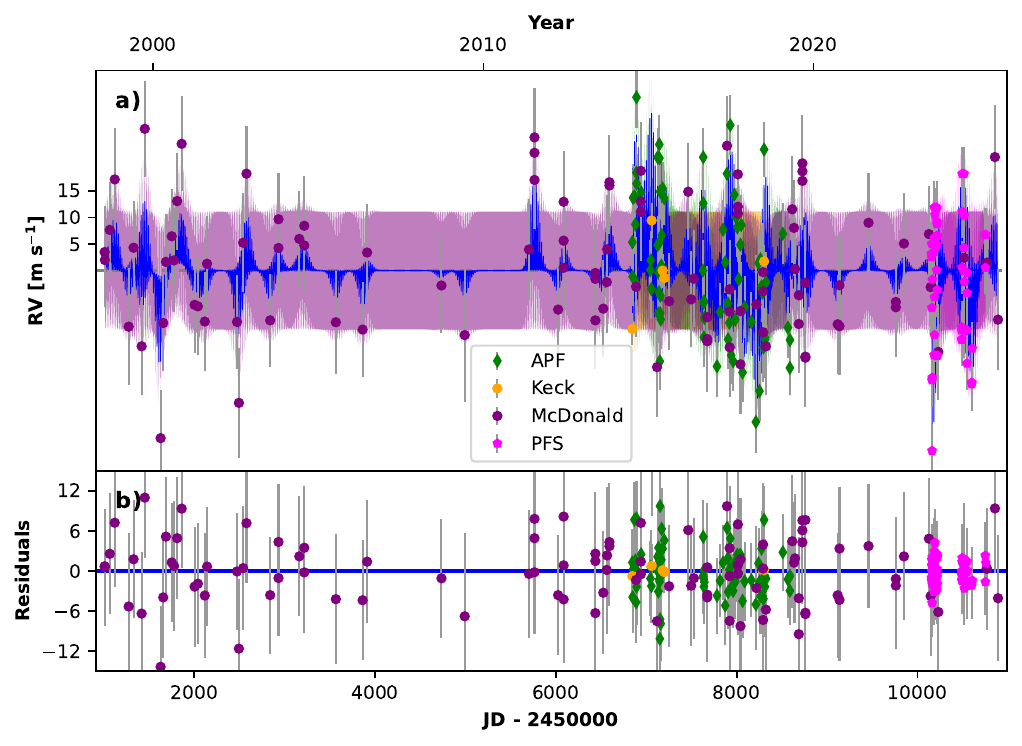}
\caption{
Quasi-periodic Gaussian process (GP) fit to the combined RVs of 70~Oph~A. This is the best-fit model. The activity model is fit simultaneously to data from APF (green diamonds), Keck/HIRES (orange circles), McDonald (purple circles), and PFS (magenta pentagons). The shaded region shows the activity predictive mean and 1$\sigma$ uncertainty. 
The quasi-periodic kernel captures rotationally modulated stellar activity with a characteristic period of $\sim19.8$~days. Panel (a) shows the RV time series and activity model; panel (b) shows the residuals after subtracting the activity model.}
\label{fig:70OphA_qp_gp}
\end{figure*}

\begin{figure*}
\centering
\includegraphics[width=\linewidth]{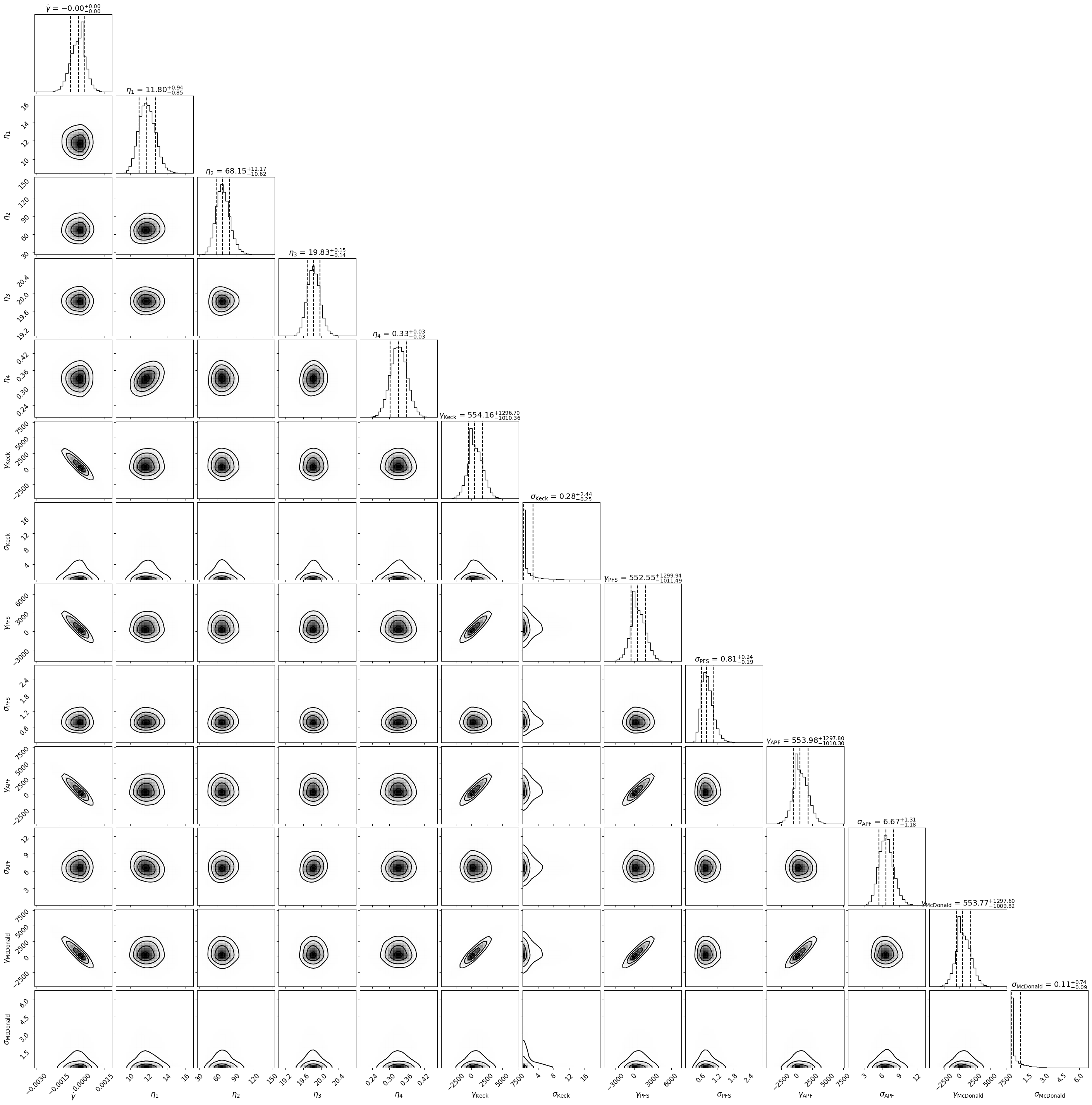}
\caption{
Posterior distributions for the quasi-periodic Gaussian process activity model fit to the RVs of 70~Oph~A. 
The GP hyperparameters include the amplitude ($\eta_1$), exponential decay timescale ($\eta_2$), periodic timescale ($\eta_3$), and coherence scale ($\eta_4$). 
Instrument-specific velocity offsets ($\gamma$) and jitter terms ($\sigma$) for Keck/HIRES, PFS, APF, and McDonald are also shown. The diagonal panels show the margin
alized posterior distributions, while the off-diagonal panels display the joint posterior correlations. 
The periodic timescale is tightly constrained at $\sim 19.8$~days, consistent with stellar rotation.}
\label{fig:70OphA_qp_corner}
\end{figure*}

\subsection{Null Model}
As a baseline for model comparison, we first fit a null model consisting
only of per-instrument radial velocity offsets ($\gamma$) and white-noise
jitter terms ($\sigma$), with no Keplerian signals and no Gaussian Process
activity component. This null model yields ${\rm BIC} = 1890.89$ (Table~\ref{tab:model_comparison}) and serves as the reference model against which we assess whether the
additional complexity of the Keplerian-only, activity-only, and hybrid models is statistically justified.

\subsection{Keplerian-Only Models}

We then test whether the dominant periodicities near 20, 10, and $\sim$7~days identified in the periodogram of 70~Oph~A can be explained by a compact multi-planet system. We assume circular Keplerian orbits ($e=0$, $\omega=0$), with independent velocity offsets ($\gamma$), white-noise jitter terms for each instrument, and a linear trend. Orbital periods are initialized near the periodogram peaks and fit using \texttt{RadVel}, allowing all periods, semi-amplitudes, times of conjunction, instrument offsets, jitter terms, and the linear slope to vary freely. Uniform priors are adopted for orbital periods and phases, semi-amplitudes are restricted to be non-negative, and jitter terms are assigned log-uniform priors. Posterior distributions are sampled using an affine-invariant MCMC ensemble with 50 walkers and 50,000 steps, discarding burn-in after convergence is achieved. All parameters satisfy $\hat{R}<1.01$, with effective sample sizes exceeding $10^3$ samples.

The two-Keplerian model converges to periods of $10.87^{+0.18}_{-0.26}$ and $20.32^{+0.01}_{-0.28}$~days, with semi-amplitudes that are weakly constrained ($K_1=-0.02^{+2.75}_{-2.61}$~m\,s$^{-1}$; $K_2=4.94^{+2.23}_{-10.92}$~m\,s$^{-1}$). This model yields $\ln\mathcal{L}=-885.86$, ${\rm BIC}=1847.85$, and an RMS of 11.59~m\,s$^{-1}$. When we add a third circular Keplerian component, the likelihood improves only marginally ($\ln\mathcal{L}=-882.04$), while the Bayesian Information Criterion increases to ${\rm BIC}=1856.53$. The additional free parameters are therefore not justified, and the three-Keplerian model is disfavored.

\subsection{Activity-Only Model}

We next model the residual RV variability using a quasi-periodic (QP) activity model. The covariance kernel is parameterized by an amplitude $A$, an exponential decay timescale $\tau$, a periodic timescale $P_{\rm rot}$, and a coherence scale $w$. Based on prior estimates of the stellar rotation period, we initialize the periodic timescale near $P_{\rm rot}\sim20$~days and allow it to vary freely in the fit. We ran MCMC for $10^5$ steps per chain and three independent ensembles until convergence is reached. 

%Even though we subtracted the binary orbit, we still include small velocity offsets and white-noise jitter terms for each instrument. We note that because the signals of interest have periods much shorter than the binary period, subtracting the binary trend has little effect on the analysis. One caveat of our fit is that we allowed the RV semi-amplitude to vary symmetrically about zero in the sampling. Negative values are degenerate with a phase shift of $\pi$ and correspond to the same physical solution as positive values. In practice, this choice does not affect our inferred constraints. We ran MCMC for $10^5$ steps per chain and three independent ensembles until convergence is reached. 

The activity-only fit to the binary-subtracted radial velocities is shown in Figure~\ref{fig:70OphA_qp_gp}, and the posterior distributions of the fitted parameters are shown in Figure~\ref{fig:70OphA_qp_corner}. The activity-only model yields $\ln\mathcal{L}=-863.34$, ${\rm BIC}=1797.37$, and an RMS of 4.11~m\,s$^{-1}$, improving upon the Keplerian-only fits. The inferred parameters for the activity model are 
$A = 11.87^{+1.00}_{-0.91}$~m\,s$^{-1}$, 
$\tau = 72^{+18}_{-14}$~days, and 
$P_{\rm rot} = 19.76^{+0.19}_{-0.16}$~days. The periodic timescale is tightly constrained and consistent with rotationally modulated stellar activity. The coherence scale $w = 0.33^{+0.03}_{-0.03}$ indicates moderately stable active regions evolving over several rotation cycles. 

\subsection{Hybrid Activity + Keplerian Models}

Having modeled the stellar variability with a quasi-periodic Gaussian Process activity model, we next test whether any coherent planetary signals remain after accounting for stellar activity. We fit hybrid models with an activity component and a single Keplerian component at candidate periods of 7, 10, 12, and 60~days.

For the 7-day signal, the hybrid model yields $P = 7.21^{+0.38}_{-0.36}$~days with a semi-amplitude of $K = 0.3^{+2.1}_{-2.4}$~m\,s$^{-1}$. For the 10-day signal, the fit converges to $P = 9.83^{+0.13}_{-0.18}$~days and $K = 0.9^{+3.4}_{-4.3}$~m\,s$^{-1}$. For the 12-day signal, we obtain $P = 12.05^{+0.42}_{-0.20}$~days and $K = 0.8^{+1.9}_{-2.8}$~m\,s$^{-1}$. Finally, the 60-day candidate yields a broad posterior of $P = 60^{+55}_{-32}$~days with $K = 0.3^{+2.6}_{-2.7}$~m\,s$^{-1}$. In all cases, the activity hyperparameters are consistent with the activity-only solution, with the rotation period detected at $P_{\rm rot}\approx19.86$~days, and similar amplitudes and evolutionary timescales.

\subsection{Model Comparison}

We compare models using the Bayesian Information Criterion (BIC), which penalizes additional free parameters and approximates differences in Bayesian evidence ($\Delta \ln Z \approx -\frac{1}{2}\Delta {\rm BIC}$ for large $N$). The results are summarized in Table~\ref{tab:model_comparison}.

Relative to the null model (${\rm BIC}=1890.89$), every model that includes Keplerian signals, activity, or both yields a substantially lower BIC, indicating that some periodic or quasi-periodic structure is required. The two- and three-Keplerian models reduce the BIC by $43.0$ and $34.4$, respectively, while the activity-only model reduces it by $93.5$, the largest improvement overall. These values constitute very strong evidence that the additional complexity is warranted \citep{Kass1995}.

The activity-only model yields the lowest BIC (${\,\rm BIC}=1797.37$) and an RMS residual of 4.11~m\,s$^{-1}$. All alternative models are disfavored relative to this solution. The two Keplerian and three Keplerian models have $\Delta{\rm BIC}=50.5$ and 59.2 relative to the activity-only model, corresponding to $\Delta\ln Z \approx -25$ and $-30$, which suggests that the data is better fit by activity.

The hybrid activity plus Keplerian models likewise increase the BIC. The 10 day and 12 day models yield ${\rm BIC}=1817.53$ and 1818.96, corresponding to $\Delta{\rm BIC}=20.2$ and 21.6, respectively. The 7 day model is more strongly penalized with ${\rm BIC}=1848.20$ and $\Delta{\rm BIC}=50.8$. The 60 day model is strongly disfavored with ${\rm BIC}=1872.03$ and $\Delta{\rm BIC}=74.7$. Under conventional metric, $\Delta{\rm BIC} > 10$ means very strong evidence against the more complex model.

We therefore conclude that the activity-only model provides the statistically preferred description of the RV variability. The dominant periodicity near 20 days is consistent with stellar rotation, and no additional Keplerian signals are supported once stellar activity is modeled.

\begin{table*}[t]
\centering
\setlength{\tabcolsep}{2pt}
\caption{Model comparison for 70~Oph~A RVs after subtraction of the best-fit binary orbit.
Circular Keplerian models are assumed ($e=0$).
Values are posterior medians with 16th–84th percentile uncertainties.
}
\label{tab:model_comparison}
\begin{tabular}{lccccccccc}
\hline\hline
Parameter & Unit
& Null
%maybe change to 10, 12, 20, 60d?
& 2-Keplerian
& 3-Keplerian
& Activity
& Activity+7d
& Activity+10d
& Activity+12d
& Activity+60d \\
\hline

\multicolumn{10}{c}{\textbf{Planet Parameters}} \\[2pt]
\hline
$P_1$ & days
& ---
& $10.87^{+0.18}_{-0.26}$
& $6.58^{+0.00}_{-0.01}$
& --- 
& $7.21^{+0.38}_{-0.36}$
& $9.83^{+0.13}_{-0.18}$
& $12.05^{+0.42}_{-0.20}$
& $60^{+55}_{-32}$ \\

$T_{c,1}$ & days
& ---
& 1085.85
& 57.93
& ---
& $2471924^{+1200}_{-510}$
& $2515329^{+350}_{-510}$
& $2528363^{+1700}_{-1700}$
& $2449719^{+5000}_{-6000}$ \\

$K_1$ & m\,s$^{-1}$
& ---
& $-0.02^{+2.75}_{-2.61}$
& $1.95^{+1.87}_{-4.17}$
& ---
& $0.3^{+2.1}_{-2.4}$
& $0.9^{+3.4}_{-4.3}$
& $0.8^{+1.9}_{-2.8}$
& $0.3^{+2.6}_{-2.7}$ \\[4pt]

$P_2$ & days
& ---
& $20.32^{+0.01}_{-0.28}$
& $13.23^{+0.00}_{-0.00}$
& ---
& --- & --- & --- & --- \\

$T_{c,2}$ & days
& ---
& 244.12
& -6.10
& ---
& --- & --- & --- & --- \\

$K_2$ & m\,s$^{-1}$
& ---
& $4.94^{+2.23}_{-10.92}$
& $3.10^{+1.50}_{-4.20}$
& ---
& --- & --- & --- & --- \\[4pt]

$P_3$ & days
& ---
& ---
& $20.21^{+0.11}_{-0.18}$
& ---
& --- & --- & --- & --- \\

$T_{c,3}$ & days
& ---
& ---
& 41.36
& ---
& --- & --- & --- & --- \\

$K_3$ & m\,s$^{-1}$
& ---
& ---
& $3.95^{+2.52}_{-9.16}$
& ---
& --- & --- & --- & --- \\[6pt]

\hline
\multicolumn{10}{c}{\textbf{Activity Hyperparameters}} \\[2pt]
\hline
$A$ & m\,s$^{-1}$
& ---
& ---
& ---
& $11.80^{+0.94}_{-0.85}$
& $11.8^{+1.0}_{-0.9}$
& $11.71^{+1.05}_{-0.91}$
& $11.85^{+1.09}_{-0.89}$
& $11.83^{+1.03}_{-0.90}$ \\

$\tau$ & days
& \textcolor{black}{---}
& ---
& ---
& $68.15^{+12.22}_{-10.68}$
& $68^{+17}_{-14}$
& $67.76^{+16.71}_{-13.54}$
& $73.99^{+19.78}_{-15.08}$
& $70.57^{+16.92}_{-13.31}$ \\

$P_{\rm rot}$ & days
& \textcolor{black}{---}
& ---
& ---
& $19.83^{+0.15}_{-0.14}$
& $19.86^{+0.18}_{-0.18}$
& $19.86^{+0.18}_{-0.18}$
& $19.86^{+0.15}_{-0.16}$
& $19.85^{+0.17}_{-0.15}$ \\[6pt]

\hline
\multicolumn{10}{c}{\textbf{Instrument Parameters}} \\[2pt]
\hline
$\gamma_{\rm HIRES}$ & m\,s$^{-1}$
& \textcolor{black}{$623^{+1100}_{-670}$}
& $-1.55$
& $-0.83$
& $554^{+1302}_{-1016}$
& $2.7^{+429}_{-321}$
& $1.02^{+72.86}_{-39.06}$
& $1.97^{+551.44}_{-144.86}$
& $2.72^{+771.26}_{-347.04}$ \\

$\sigma_{\rm HIRES}$ & m\,s$^{-1}$
& \textcolor{black}{$7.8^{+4.0}_{-2.2}$}
& $6.03$
& $4.42$
& $0.3^{+2.5}_{-0.25}$
& $0.41^{+5.73}_{-0.38}$
& $0.53^{+5.81}_{-0.49}$
& $0.52^{+5.46}_{-0.48}$
& $0.58^{+6.01}_{-0.54}$ \\

$\gamma_{\rm PFS}$ & m\,s$^{-1}$
& \textcolor{black}{$624^{+1100}_{-670}$}
& $0.56$
& $-0.10$
& $553^{+1306}_{-1017}$
& $1.65^{+430}_{-322}$
& $-1.14^{+69.96}_{-43.45}$
& $2.87^{+550.66}_{-143.41}$
& $0.93^{+771.63}_{-348.27}$ \\

$\sigma_{\rm PFS}$ & m\,s$^{-1}$
& \textcolor{black}{$11.8^{+1.4}_{-1.1}$}
& $10.21$
& $10.67$
& $0.81^{+0.25}_{-0.19}$
& $0.85^{+0.29}_{-0.22}$
& $0.84^{+0.33}_{-0.21}$
& $0.82^{+0.31}_{-0.19}$
& $0.82^{+0.31}_{-0.20}$ \\

$\gamma_{\rm APF}$ & m\,s$^{-1}$
& \textcolor{black}{$624^{+1100}_{-670}$}
& $0.13$
& $0.22$
& $554^{+1302}_{-1015}$
& $1.73^{+432}_{-321}$
& $0.71^{+70.26}_{-42.80}$
& $2.46^{+552.25}_{-146.73}$
& $2.25^{+771.24}_{-345.63}$ \\

$\sigma_{\rm APF}$ & m\,s$^{-1}$
& \textcolor{black}{$13.5^{+1.3}_{-1.1}$}
& $12.61$
& $12.76$
& $6.7^{+1.3}_{-1.2}$
& $6.76^{+1.67}_{-1.29}$
& $6.82^{+1.62}_{-1.35}$
& $6.77^{+1.52}_{-1.32}$
& $6.95^{+1.55}_{-1.44}$ \\

$\gamma_{\rm McD}$ & m\,s$^{-1}$
& \textcolor{black}{$623^{+1100}_{-670}$}
& $1.03$
& $-0.77$
& $554^{+1301}_{-1015}$
& $1.57^{+429}_{-319}$
& $0.60^{+71.72}_{-42.53}$
& $2.26^{+551.73}_{-145.48}$
& $2.33^{+770.55}_{-346.19}$ \\

$\sigma_{\rm McD}$ & m\,s$^{-1}$
& \textcolor{black}{$11.07^{+1.0}_{-0.89}$}
& $10.22$
& $9.58$
& $0.11^{+0.75}_{-0.091}$
& $0.21^{+1.25}_{-0.18}$
& $0.27^{+1.48}_{-0.24}$
& $0.23^{+1.57}_{-0.20}$
& $0.19^{+1.41}_{-0.17}$ \\

Slope & m\,s$^{-1}$\,day$^{-1}$
& \textcolor{black}{$-2.5\times10^{-4}$}
& $-3.6\times10^{-4}$
& $-3.7\times10^{-4}$
& $-2.3\times10^{-4}$
& $0.02\times10^{-4}$
& $2\times10^{-7}$
& $-2\times10^{-6}$
& $-1\times10^{-6}$ \\

RMS & m\,s$^{-1}$
& \textcolor{black}{12.47}
& 11.59
& 11.49
& 4.107
& 4.09
& 3.658
& 4.107
& 4.340 \\[4pt]

\hline
\multicolumn{10}{c}{\textbf{Model Statistics}} \\[2pt]
\hline
$\ln \mathcal{L}$ & ---
& \textcolor{black}{$-920.971$}
& $-885.86$
& $-882.04$
& $-863.336$
& $-880.60$
& $-865.260$
& $-865.974$
& $-892.509$ \\

$N_{\rm RV}$ & ---
& \textcolor{black}{$230$}
& 230
& 230
& 230
& 230
& 230
& 230
& 230 \\

$k$ & ---
& \textcolor{black}{$9$}
& 14
& 17
& 13
& 16
& 16
& 16
& 16 \\

BIC & ---
& \textcolor{black}{$1890.89$}
& 1847.85
& 1856.53
& 1797.37
& 1848.20
& 1817.53
& 1818.96
& 1872.03 \\
\hline
\hline
\end{tabular}
\begin{flushleft}
\small \textcolor{black}{$^a$ For the Keplerian models, $T_c$ values are reported relative to a reference epoch of BJD $= 2457920.804$, corresponding to the median time of the RV dataset used in the fit. For the activity-Keplerian hybrid models, $T_c$ values are reported in absolute BJD.}
\end{flushleft}
\end{table*}

\section{Injection and Recovery Tests for 70~Oph~AB}
\label{sec:injection-recovery}

The long baseline of RV data enables us to derive sensitivity curves and exclude or rule out planetary companions of certain masses. To place upper limits on our detection, we performed injection--recovery experiments using the publicly available \texttt{rvsearch} pipeline developed by the California Planet Search (CPS) collaboration \citep{Rosenthal_2021}. The \texttt{rvsearch} framework builds upon the \texttt{RadVel} Keplerian modeling package \citep{Fulton_2018} and provides a pipeline for blind planet searches and completeness calculations. 

For 70~Oph~A, we utilize the combined dataset from Keck/HIRES, APF, McDonald and PFS spanning 27 years as we did for the GPR analysis. For 70~Oph~B, the analysis is restricted to the 2-year high-precision PFS dataset, which is the only RV dataset available.

Synthetic Keplerian signals were injected directly into the observed radial velocity time series using the CPS injection module. For each trial, a circular orbit ($e=0$) was assumed:

\begin{equation}
    V_r(t) = K \sin\left( \frac{2\pi t}{P} + \phi \right),
\end{equation}

where $K$ is the radial velocity semi-amplitude, $P$ is the orbital period, and $\phi$ is the orbital phase.

The injected periods were sampled logarithmically over a range extending from short periods (1--5 days) up to $\sim 0.9$ times the full temporal baseline of the dataset (27 years for 70~Oph~A). Velocity semi-amplitudes were sampled over a grid spanning sub-meter-per-second to tens of meters per second, and multiple orbital phases were injected at each grid point to marginalize over phase dependence.

\begin{figure*}[t]
    \centering
    \begin{tabular}{cc}
        \includegraphics[width=0.5\textwidth]{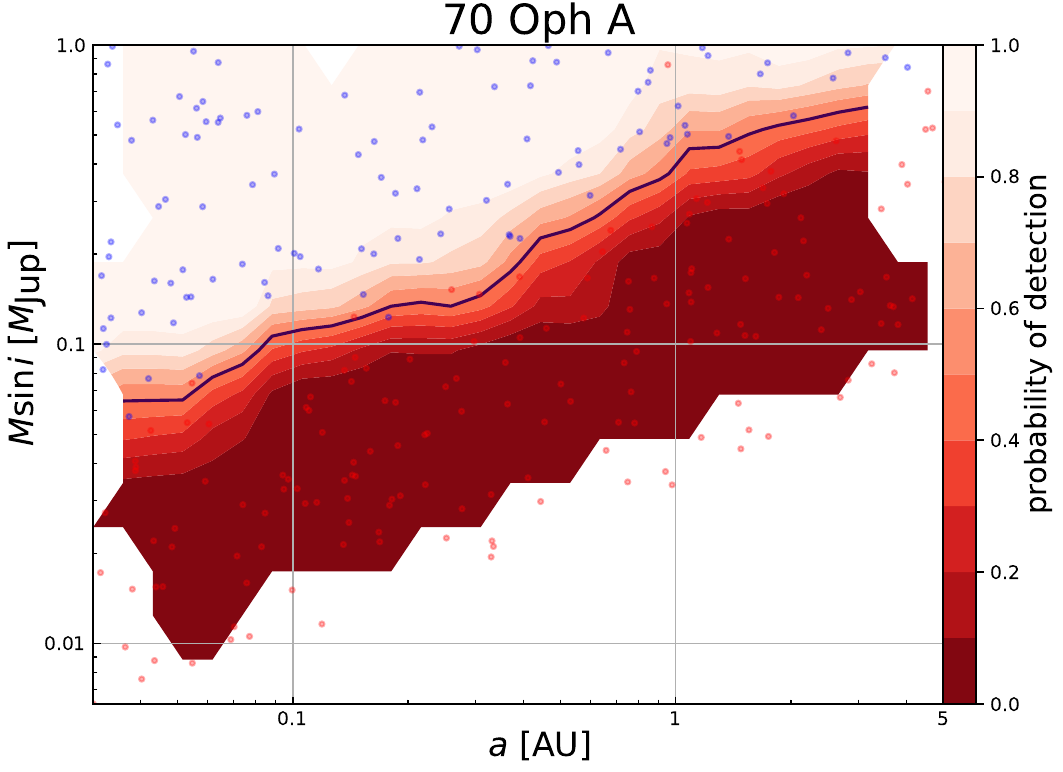} & 
        \includegraphics[width=0.5\textwidth]{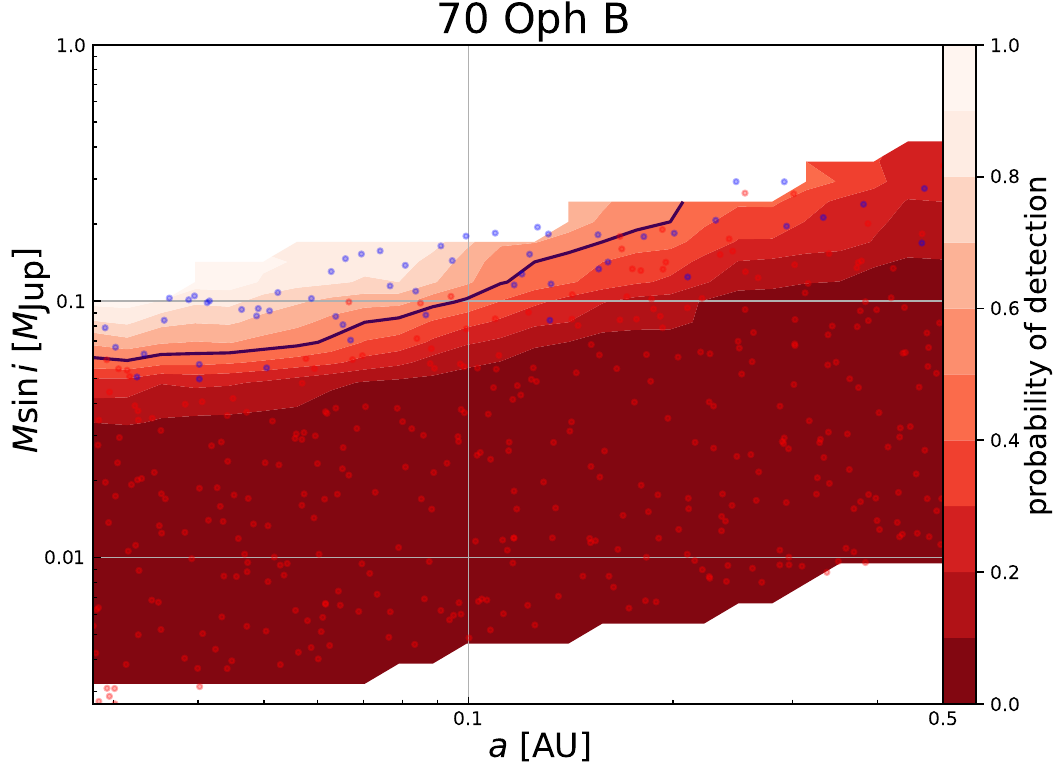}
    \end{tabular}
    
    \caption{Detection completeness for 70 Oph A (left) and 70 Oph B (right) derived using \texttt{rvsearch} injection--recovery simulations. Synthetic Keplerian signals were injected into the radial-velocity data assuming circular orbits ($e=0$). The completeness map for 70 Oph A uses the combined Keck/HIRES, APF, and PFS dataset, while the map for 70 Oph B is based on the high-cadence PFS data alone. Background shading indicates the interpolated recovery probability as a function of semi-major axis and $M\sin i$. Individual injection trials are shown as points, with blue circles denoting recovered signals and red circles indicating non-recovered signals. 
    The solid black curve marks the 50\% completeness contour, and the other gradient contour lines show completeness levels from 10\% to 90\%.}
    \label{fig:completeness_comparison}
\end{figure*}

For each injected signal, \texttt{rvsearch} performs an uninformed sequential periodogram search and Keplerian fitting procedure using the same GP modeling in the previous section. A signal is considered successfully recovered if the highest-likelihood model identifies the injected period within a specified tolerance and exceeds the adopted false-alarm probability threshold.

Because injections are performed in velocity semi-amplitude space, the completeness limits apply directly to $M_p \sin i$ and do not require any assumption about orbital inclination or coplanarity with the binary plane. The resulting completeness maps therefore quantify sensitivity purely as a function of observable Doppler amplitude given the measured sampling and noise properties of the data. For each grid point in $(P, K)$ space, the detection fraction was computed as the fraction of injected phases successfully recovered. These detection fractions were used to construct two-dimensional completeness maps in semi-major axis and minimum mass.

The recovery statistics were used to compute a two-dimensional completeness surface in semi-major axis $a$ and projected mass $M_p \sin i$. Figure~\ref{fig:completeness_comparison} presents the resulting completeness contours for both stars. We show the 50\% detection probability levels, above which we can rule out the presence of planetary companions. For 70~Oph~A, we exclude $M\sin i \gtrsim 0.3\,M_{\rm Jup}$ at 1~AU and $\gtrsim 0.5\,M_{\rm Jup}$ at 2~AU, ruling out Jupiter-mass planets throughout the inner $\sim5$~AU. For 70~Oph~B, sensitivity extends to $\sim0.5$~AU, where we exclude $M\sin i \gtrsim 0.25$--$0.3\,M_{\rm Jup}$.

\section{Dynamical Limits on Planetary Companions}
\label{sec:limits}

The presence of the binary star in the 70~Oph system can place significant constraints upon the presence of possible planetary orbits around the primary star. In particular, this may inhibit planetary orbits at important locations for planetary evolution, such as beyond the snow line or within the HZ \citep{kane2025b}. The HZ is the boundary wherein liquid water could exist on the surface of an exoplanet, given it has appropriate levels of atmospheric pressure and greenhouse warming \citep{Kasting1993}. The HZ boundaries are a useful tool for determining target selections for follow-up missions that aim to characterize potentially habitable environments \citep{kane2012}. In addition to their gravitationally perturbative effect, the presence of a binary companion may influence the HZ boundaries, depending on whether the binary orbit is of a P-type \citep{haghighipour2013c,kane2013a,cukier2019} or an S-type \citep{eggl2012,kaltenegger2013b,wang2017d} configuration. Our dynamical simulations are assessing the case of a planet in an S-type orbit around the primary star. Based on the binary orbital parameters provided in Table~\ref{tab:orbit_results}, the closest approach of the stellar components is $\sim$11.6~AU, which is larger than Saturn's semi-major axis. The relatively low mass of the secondary star, in conjunction with its distance at periastron, suggests that the outer boundaries of the primary star's HZ experience only minor modifications due to the effects of the secondary.

To calculate the HZ boundaries of 70~Oph~A, we adopt the methodology described by \citet{Kopparapu2013,Kopparapu2014} and the stellar properties listed in Table~\ref{tab:stellar_params}. These calculations establish the limits on both the conservative HZ (CHZ) and optimistic HZ (OHZ), based on the runaway and maximum greenhouse limits and empirical evidence of the history of surface liquid water on Venus and Mars, respectively \citep{Kane2016,Kane2024}. The bounds on the CHZ and OHZ are 0.711--1.273~AU and 0.561--1.342~AU, respectively. Combining these HZ boundaries with the analysis of 70 Oph B's dynamical effects will reveal whether undetected exoplanets could be present within the HZ of 70~Oph~A. 

A first-order estimate for the dynamical limits on planetary orbits imposed by the binary companion may be derived using the empirically derived formula of \citet{Holman1999}, which provides the  critical semi-major axis for a circumstellar orbit, beyond which the orbit is likely to be unstable. Using the binary star properties for 70~Oph described in Table~\ref{tab:orbit_results}, we calculate a critical semi-major axis of 2.88~AU.

To assess the viability of planetary orbits around the primary star, we carried out a series of dynamical simulations using the REBOUND N-body integrator package \citep{rein2012}, which applies the symplectic integrator WHFast \citep{rein2015}. Each simulation tracked the orbital evolution of an Earth-mass planet injected into a circular orbit that is coplanar with the orbit of the binary. We explored the semimajor axis range of 0.5--10.5~AU, divided into 2000 steps of 0.005~AU. \textcolor{black}{Each simulation used a single test particle per semi-major axis step, initialized with a fixed argument of pericenter ($\omega = 90 ^\circ$) and a true anomaly defaulting to $0 ^\circ$. Only prograde orbits were considered in these simulations.} We used an integration time of $10^7$~yr per simulation, based on the results of \citet{rabl1988}, which determined that an integration time equivalent of $\sim$300 binary orbits is sufficient to adequately sample the dynamical stability of other orbits within the system. Orbital elements were recorded every 100~yrs of simulation time, and with a time step of 5 days to ensure adequate time resolution of planetary perturbations due to the binary companion. At the conclusion of each simulation, we calculated the maximum eccentricity of the planet and the percentage of the integration for which the injected planet remained within the system. Loss of the planet generally means that it has either been ejected from the system or consumed by the gravitational well of the primary star.

\begin{figure*}
  \begin{center}
    \includegraphics[width=16.05cm]{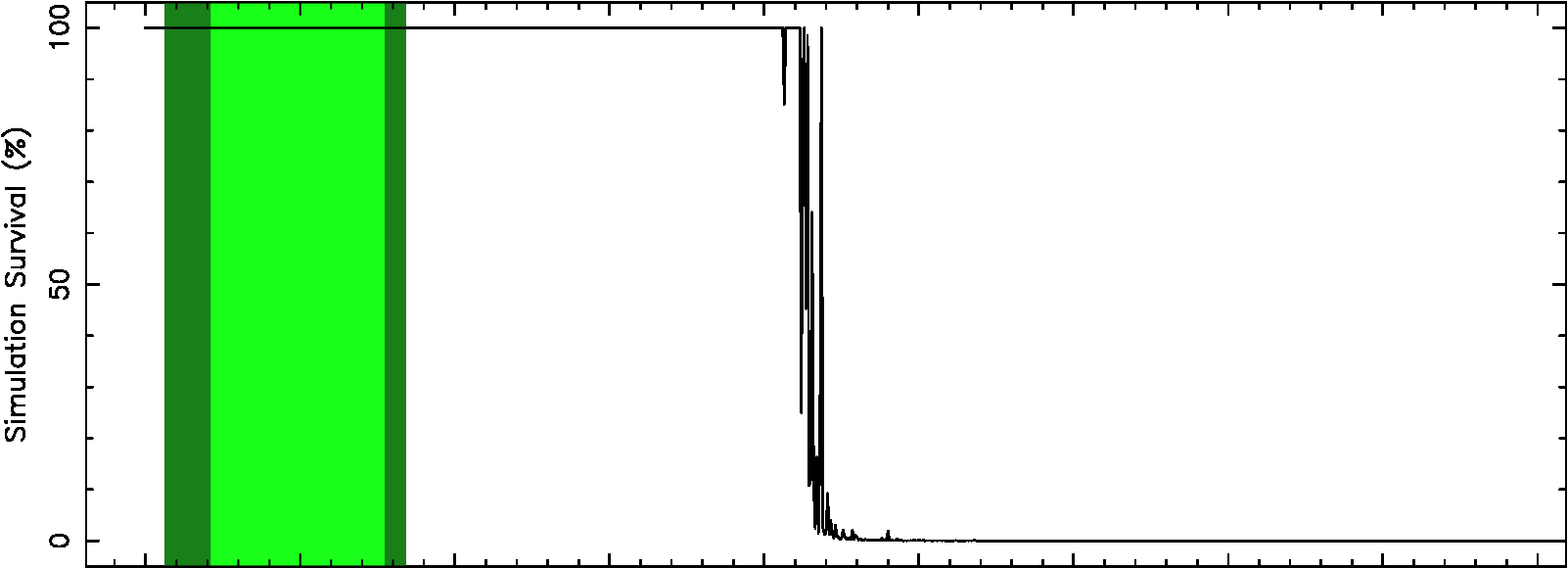} \\
    \includegraphics[width=16.0cm]{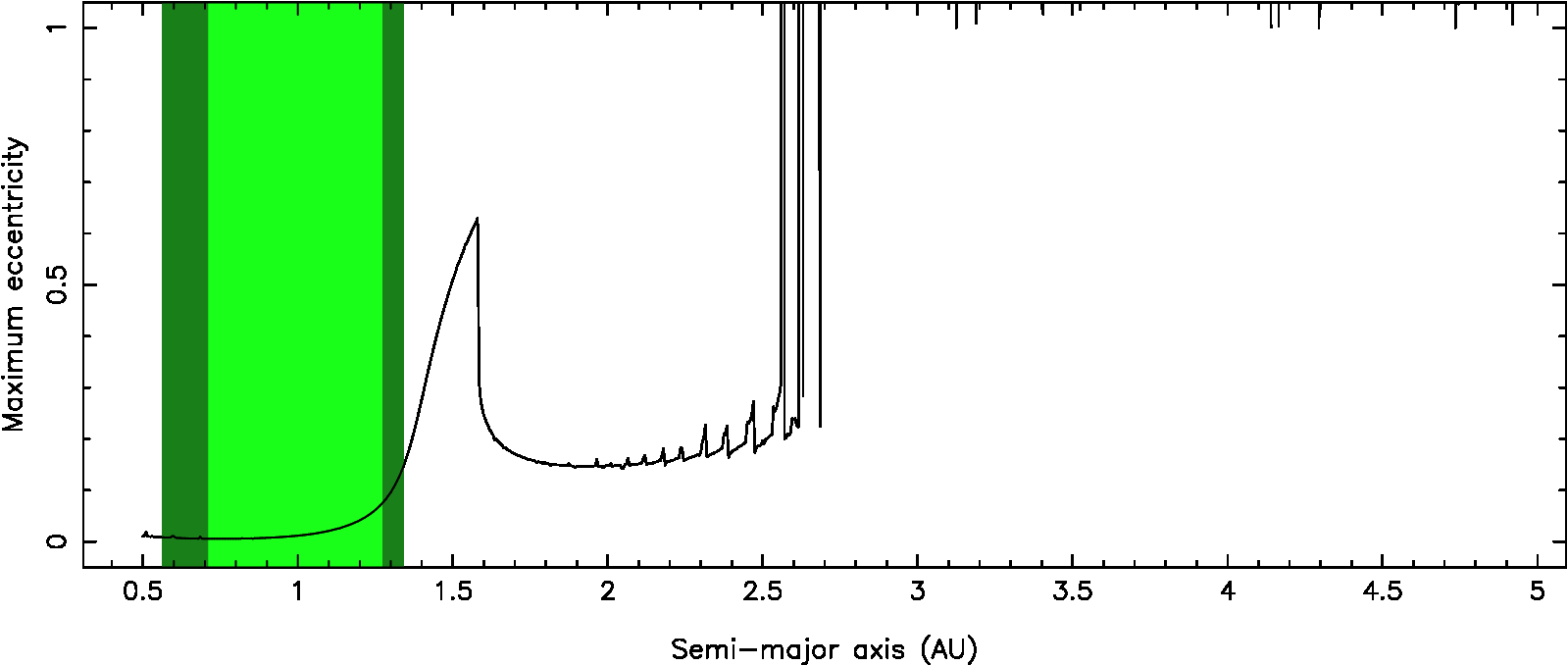}
  \end{center}
  \caption{Results of the dynamical simulation for the 70~Oph system as a function of semi-major axis of the injected planet. The top panel provides the percentage survival calculations for the planet, and the bottom panel shows the maximum eccentricity of the planet. The extent of the HZ is shown in green, where the light green and dark green regions indicate the CHZ and OHZ, respectively.}
  \label{fig:sim}
\end{figure*}

The results of our dynamical simulations are shown in Figure~\ref{fig:sim} as a function of the injected planet's semi-major axis from the primary star. The top panel shows the percentage of simulations in which the planet survives within the system at each semi-major axis. The bottom panel shows the maximum eccentricity of the planetary orbit during each simulation. For both panels, the light green regions indicates the extent of the CHZ, and the dark green shows the additional of the OHZ. The results exclude the presence of any planetary orbits beyond $\sim$2.5~AU from the primary star due to their dynamical instability. Within the HZ, orbits may remain in a stable configuration with only moderate induced eccentricity due to the secondary star. Between the outer edge of the HZ and the dynamical limit at $\sim$2.5~AU, there are numerous minor peaks, which are the result of mean-motion resonance (MMR) locations with the orbit of the binary \citep{raymond2008b,kane2023a}. In particular, a large peak in eccentricity occurs around 1.5~AU from the primary, and with eccentricities reaching as high as 0.63. Such large orbital perturbations may be produced by the interaction of the orbital precession of the planet with the resonance locations of the binary \citep{touma2015}. Overall, our results demonstrate that it is possible for the 70 Oph system to host planets on S-type orbits around the primary star, including terrestrial planets on stable orbits within the HZ region.

%The analysis of the GLS periodograms shows that the phase of the RV signal (and in particular the peak in its value) is about xx degrees behind the one observed in FWHM and log R'HK, respectively. This lag is expected if the RV signal is caused by stellar spot (cite Forveille 2009). Indeed, when the active regions appear and occupy the blueshifted side of the star, the RV will show an increasing value with time. Simulation with the SOAP code (Boisse 2011) show that the maximum of this RV will occur when the spots are xx degrees from meridian. This is consistent with the observed lag. The value will decrease to zero when the spot is at meridian. This instant sets the highest activity level as the active region shows its maximum projected area.

\section{Discussion}
\label{sec:discussions}

% \subsection{System Age}

% The precise dynamical masses and well-constrained stellar radii derived in this work allow us to estimate the surface gravity, $\log g$, for both components and place constraints on the ages. From the measured masses, $M_{\rm A} = \texttt{[value]}\,M_\odot$ and $M_{\rm B} = \texttt{[value]}\,M_\odot$, and radii, $R_{\rm A} = \texttt{[value]}\,R_\odot$ and $R_{\rm B} = \texttt{[value]}\,R_\odot$, we compute empirical surface gravities of $\log g_{\rm A} = \texttt{[value]}$ and $\log g_{\rm B} = \texttt{[value]}$. Comparison with stellar evolutionary models such as \texttt{[MIST/PARSEC/Dartmouth]} then provides an independent age estimate for the system and may help break the degeneracy in ages discussed in Section~\ref{sec:stellarproperty}.

\subsection{Implications for Habitability}

With dynamical masses measured to $<1\%$ precision and an orbit constrained by more than a century of astrometry and radial velocities, 70~Oph~AB is now among the best-characterized nearby solar-type binaries. This is especially valuable for habitability studies in binary stars, where even modest changes in eccentricity or periastron distance can substantially modify the extent of long-term stable zones. In this context, 70~Oph~AB is interesting because its mass ratio and orbital separation are similar to those of $\alpha$~Cen~AB. Dynamical studies of $\alpha$~Cen~AB have shown that binary perturbations can excite orbital inclinations and, over long timescales, torque planets away from strict coplanarity with the binary plane \citep{Quarles_2016}.
 
Our results are consistent with \citet{Jankowski_2025}, who recently studied 70~Oph~A/B using $N$-body simulations and found that the habitable zones around both stars are dynamically viable in the coplanar case, but are destabilized in strongly inclined configurations by Kozai--Lidov oscillations. Adopting earlier literature orbits, they derived habitable zones of 0.52--1.25 AU for 70~Oph~A and 0.30--0.71 AU for 70~Oph~B, along with an analytic critical S-type stability limit of $a_c \simeq 3.03$ AU. Our simulation yields a slightly smaller stable circumprimary region ($\sim$2.5--2.9 AU) for 70~Oph~A, but leads to the same conclusion that the habitable zone is dynamically viable provided planetary orbits are roughly aligned with the binary plane.

Past studies placed upper limits on planetary companions in the 70~Oph system, but our analysis benefits from the longest RV baseline yet assembled for this binary. Using only McDonald Observatory velocities for 70~Oph~A, \citet{Wittenmyer2006} ruled out companions at the 99\% confidence level with $M\sin i \sim 1\,M_{\rm Jup}$ at 1 AU, $\sim3\,M_{\rm Jup}$ at 2 AU, and $\sim6\,M_{\rm Jup}$ at 3 AU. With the expanded dataset analyzed here, we place upper limits on planets around both 70~Oph~A and 70~Oph~B, albeit with weaker constraints for 70~Oph~B because of the more limited RV coverage. In the habitable zone of 70~Oph~A, at $\sim$2.5 AU, our 90\% confidence limit rules out companions with $M \gtrsim 1\,M_{\rm Jup}$, while in the habitable zone of 70~Oph~B, at $\sim$0.5 AU, we can also rule out companions more massive than roughly Jupiter. The much longer baseline and expanded RV dataset now allow substantially stronger empirical constraints on planets in the habitable zone of 70~Oph~A. Our results also complement nicely the recent work of \citet{Hirsch_2021} and \citet{Harada_2025}. \citet{Hirsch_2021} analyzed California Planet Search RVs, together with HIRES and APF observations of the A component, in a nearby-star planet-occurrence survey, whereas \citet{Harada_2025} proposed 70~Oph as a potential HWO target. However, neither study was able to place meaningful upper limit constraints due to the limited amount of public RV data available at the time.

We conclude that 70~Oph~AB is capable of hosting habitable-zone planets, and our injection-recovery tests do not exclude the possibility of sub-Jupiter-mass planets in the habitable zone of 70~Oph~A or 70~Oph~B.

\subsection{No Evidence for Planets in the Current RV Data}

Despite favorable dynamical conditions, the radial velocities reveal no statistically significant planetary signals around either stellar component. We find that the RV variability of 70~Oph~A is dominated by a $\sim$20-day signal and its harmonics. This is inconsistent with any planetary companion, which would produce a coherent Keplerian signature. Instead, the harmonic structure indicates rotational modulation by non-axisymmetric surface inhomogeneities.

In magnetically active K dwarfs, starspots and plage regions perturb spectral line profiles through a combination of flux suppression and inhibition of convective blueshift. As these active regions rotate across the visible hemisphere, the resulting line asymmetries generate quasi-periodic RV variations whose amplitude and phase depend on projected area, latitude, and spot evolution \citep[e.g.,][]{Boisse_2010}. The non-sinusoidal morphology and finite coherence timescale inferred from our GPR fits are better explained by evolving active-region complexes rather than planetary candidates. The inferred activity semi-amplitude ($A \sim 12$ m\,s$^{-1}$) is consistent with spot coverage fractions of order a few percent for K dwarfs at this activity level.

\subsection{Obliquity and Mutual Inclination}

The joint astrometric and RV solution yields a binary inclination of 
$i_{\rm bin} = 121.14^\circ \pm 0.03^\circ$ (Table~\ref{tab:orbit_results}), 
corresponding to a tilt of $|180^\circ - i| \approx 58.9^\circ$ relative to the plane of the sky. Using the interferometric radius ($R_\star = 0.831\,R_\odot$) and the rotation period from our GP fit ($P_{\rm rot} = 19.76$~days), we infer an equatorial rotation velocity of
\begin{equation}
v_{\rm eq} = \frac{2\pi R_\star}{P_{\rm rot}}
           \approx 2.1~{\rm km\,s^{-1}}.
\end{equation}

High-resolution spectroscopic analyses report projected rotation velocities for 70~Oph~A in the range $v\sin i_\star \sim 1.5$--3~km~s$^{-1}$ \citep[e.g.,][]{Herrero_2012, Brewer_2016, Johnson2010, Abdurro2022}. Taking $v\sin i_\star \approx 2$~km~s$^{-1}$ and $v_{\rm eq} \approx 2.1$~km~s$^{-1}$ implies $\sin i_\star \sim 1$, suggesting that the stellar spin axis is viewed close to edge-on. For 70~Oph~B, the rotation period and published spectroscopic line-broadening values of $v\sin i_\star \sim 2$--5~km~s$^{-1}$ \citep{Glcebocki_2005,Brewer_2016,Jonsson_2020,Freckelton_2025} are too uncertain to support a robust spin-inclination inference.

The true mutual inclination between the stellar spin and the binary orbit cannot be determined because the longitude of the stellar spin axis is unknown. In general,
\begin{equation}
\cos \Phi =
\cos i_{\rm bin}\cos i_\star
+
\sin i_{\rm bin}\sin i_\star
\cos(\Omega_{\rm bin}-\Omega_\star),
\end{equation}
where $i$ and $\Omega$ denote inclination and nodal angle. While $i_{\rm bin}$ and $\Omega_{\rm bin}$ are measured from the astrometric orbit, $\Omega_\star$ is unconstrained. 

Our dynamical calculations (Section~\ref{sec:limits}) assume coplanarity and demonstrate long-term stability for S-type orbits interior to $\sim2.5$~AU of 70~Oph~A. If the stellar spin axes and the binary orbital angular momentum are aligned as observed in many wide binaries \citep{Hale_1994}, then circumstellar planets around 70~Oph~A would likely share the same orbital plane. In that case, the planetary inclination would be similar to that of the binary, and the true planetary masses would be close to their RV minimum masses ($M_p \sin i$). However, if planetary orbits are strongly misaligned with respect to the binary plane, Kozai–Lidov oscillations \citep{Kozai1962, Lidov1962} could excite large eccentricities and reduce long-term stability. 

\subsection{Direct Imaging and Astrometry Opportunity}

70~Oph~AB presents a unique opportunity for the search and characterization of habitable planets in binary systems. Like $\alpha$~Cen, whose distant tertiary companion Proxima~Cen hosts the habitable-zone terrestrial planet Proxima~b, and whose primary $\alpha$~Cen~A may host a planet \citep{Beichman_2025, Sanghi_2025}, 70~Oph is one of the nearest binaries where the habitable zone (HZ) can be spatially resolved. Because the habitable zone of 70~Oph~A lies at $\sim$0.7–1.3 AU and the system is only 5.1 pc away, the corresponding angular separation is $\sim$140–250 mas. For example, a Saturn-mass planet ($\sim0.3,M_{\rm Jup}$) at 1 AU, permitted by our RV detection limits, would produce a reflected-light contrast of order $\sim10^{-7}$ and an astrometric signature of $\sim70~\mu$as (increasing to $\sim100~\mu$as for a $0.5,M_{\rm Jup}$ companion). 

Because habitable-zone planets in 70~Oph would be difficult to detect in reflected light, Roman \citep{Fluckiger_2025} and HWO \citep{Gaudi_2020, Feinberg, Harada_2024} offer the most promising path toward achieving the required direct-imaging contrasts. The equilibrium temperature (assuming no significant internal energy for the age of the system) makes the mid-IR best for detecting thermal emission. For METIS on the 39-meter ELT, \citet{Bowens_2021} suggest that, for nearby Sun-like systems such as $\alpha$~Cen, a Saturn-mass planet at 1 AU might be detectable in contrast and sensitivity in about 1 hour of integration time. In addition, the astrometric signature of such a companion could in principle be accessible to Gaia if the challenges associated with very bright-star saturation can be overcome \citep{Sahlmann_2016}. Furthermore, future relative-astrometry missions such as SHERA \citep{Christiansen2025SHERA} are also particularly well suited to this system. For an Earth-mass planet at 1 AU, the expected astrometric signature is of order $\sim$0.2--0.3 $\mu$as, while a super-Earth (5--10 $M_\oplus$) would induce several $\mu$as. 
Given the brightness and proximity of 70~Oph, microarcsecond astrometry could probe the terrestrial-mass regime inaccessible to RV surveys. 

%At near-infrared wavelengths, a 30 m telescope has a diffraction limit of $\sim$8–15 mas and a typical coronagraph inner working angle of $\sim$15–40 mas, placing the habitable zone well outside the Inner Working Angles (IWA) of future Extremely Large Telescopes (ELTs) and the Habitable Worlds Observatory (HWO) \citep{Gaudi_2020, Feinberg, Harada_2024}. 

\section{Conclusions}
\label{sec:conclusions}

We present a comprehensive re-analysis of the nearby binary system 70~Oph~AB, combining new high-cadence, sub--m\,s$^{-1}$ radial velocities from Magellan/PFS with archival radial velocities and more than a century of relative and absolute astrometry to derive the orbital solution. Our main results are as follows:

\begin{itemize}

\item A joint fit of archival RVs, \textit{Hipparcos} and \textit{Gaia} absolute astrometry, and historical relative astrometry yields a highly precise three-dimensional orbital solution for the binary. We derive dynamical masses of $M_A = 0.88 \pm 0.004\,M_\odot$ and $M_B = 0.73 \pm 0.003\,M_\odot$, with fractional uncertainties below one percent. The orbital period ($P = 88.126 \pm 0.010$ yr), eccentricity ($e = 0.50015 \pm 0.00017$), and inclination are now tightly constrained, making 70~Oph~AB the best-calibrated nearby K-dwarf binaries.

\item After subtraction of the binary orbit, the residual RV variability of 70~Oph~A is dominated by signals at $\sim$20 days and its harmonics ($\sim$10 and $\sim$7 days), coincident with the stellar rotation period measured from chromospheric diagnostics. We detect no statistically significant planetary signals using both periodogram analysis and Gaussian Process Regression modeling.

\item We derive the RV sensitivity curves from the combined RV dataset using injection--recovery tests. For 70~Oph~A, we exclude companions more massive than $\sim0.3\,M_{\rm Jup}$ at 1~AU and $\sim0.5\,M_{\rm Jup}$ at 2~AU at the 50\% completeness level. Jupiter-mass planets interior to 1~AU would be recovered with near-unity completeness and are robustly ruled out. For 70~Oph~B, sensitivity is limited to $\sim$0.5~AU, where we exclude companions more massive than $\sim0.25$–$0.3\,M_{\rm Jup}$. Terrestrial-mass planets within the habitable zone lie below the present detection threshold set by stellar activity.

\item Dynamical simulations using the updated binary orbit show that S-type planetary orbits around 70~Oph~A are stable interior to $\sim$2.5--2.9~AU. The conservative and optimistic habitable zones (0.71--1.27~AU and 0.56--1.34~AU, respectively) lie within this stable region. Earth-mass planets on coplanar, near-circular orbits in the habitable zone are dynamically viable.

\end{itemize}

In summary, the present RV data reveal no evidence for planetary companions around either component of 70~Oph~AB, and the observed short-period variability is consistent with rotationally modulated stellar activity. Our RV upper limits rule out close-in giant planets in the inner few AU, while the terrestrial-mass regime is beyond current sensitivity. Together with its precisely determined binary orbit and dynamically viable habitable zone, 70~Oph~AB is a compelling benchmark for future habitable-planet searches in nearby binaries. Its proximity, brightness, and accurately measured stellar masses make it an especially attractive target for next-generation high-contrast imaging and microarcsecond astrometry. If stellar binaries can host Earth-like planets, 70~Oph is one of the best places to look.

\hspace*{+7mm}
\section*{Acknowledgments}
Part of this research was carried out at the Jet Propulsion Laboratory, California Institute of Technology, under a contract with the National Aeronautics and Space Administration (80NM0018D0004). We acknowledge support from the Heising-Simons Foundation through grant 2020-1699 for MIRAC-5. We gratefully acknowledge support through the allocation of observing time with the Planet Finder Spectrograph on the 6.5 m Magellan Clay Telescope at Las Campanas Observatory, Chile, during semesters 2023B through 2025B.

\section*{Data Availability}
The complete data underlying this article are available from the corresponding author upon request. All the {\it TESS} data used in this paper can be found in MAST: \dataset[10.17909/285z-xh42]{http://dx.doi.org/10.17909/285z-xh42}.


% \facilities{HST(STIS), Swift(XRT and UVOT), AAVSO, CTIO:1.3m,


% \software{astropy \citep{2013A&A...558A..33A,2018AJ....156..123A},  
%           Cloudy \citep{2013RMxAA..49..137F}, 
%           Source Extractor \citep{1996A&AS..117..393B}
%           }



\bibliography{sample701}{}
\bibliographystyle{aasjournalv7}

\FloatBarrier
\clearpage
\textcolor{black}{\appendix}
\FloatBarrier

\section{Orbital Solution for 70~Oph~B}
\label{app:orbit_B}

\begin{figure*}[!htbp]
\centering
\includegraphics[width=0.47\linewidth]{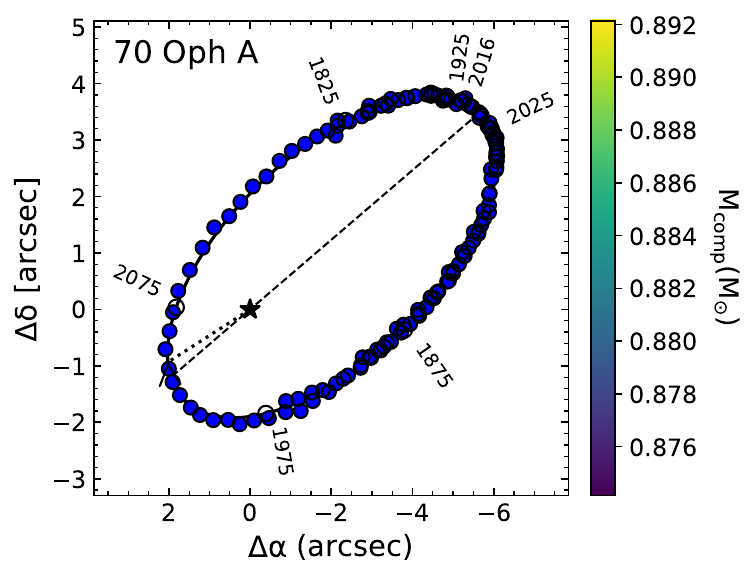}
\includegraphics[width=0.45\linewidth]{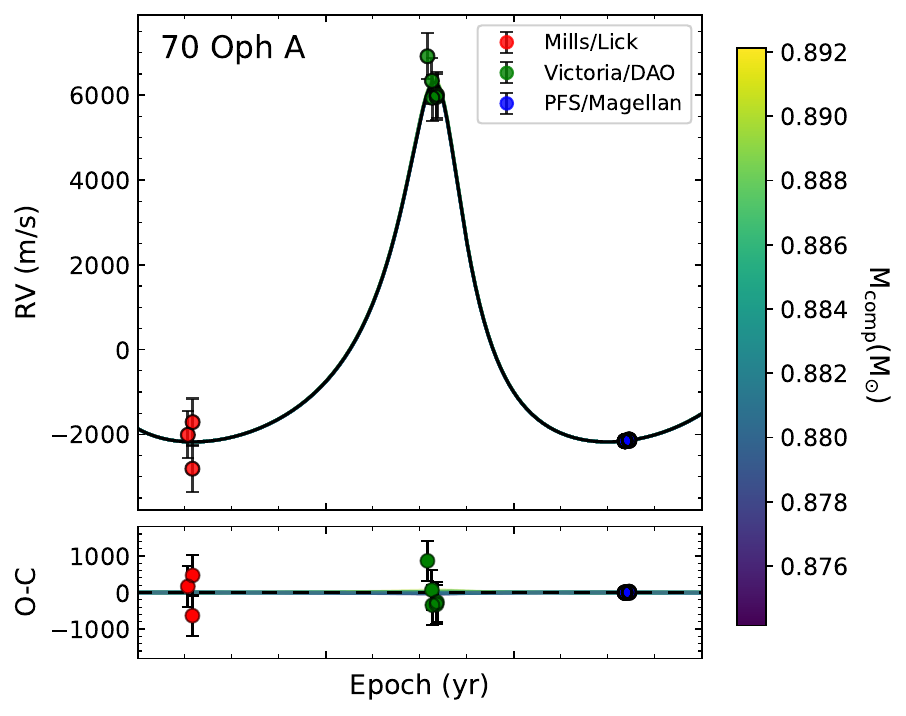}
\caption{Same as Figure~\ref{fig:binary_orbit}, but for 70~Oph~A's orbit around 70~Oph~B.}
\label{fig:binary_orbit_b}
\end{figure*}

\begin{figure*}[!htbp]
\centering
\includegraphics[width=0.45\linewidth]{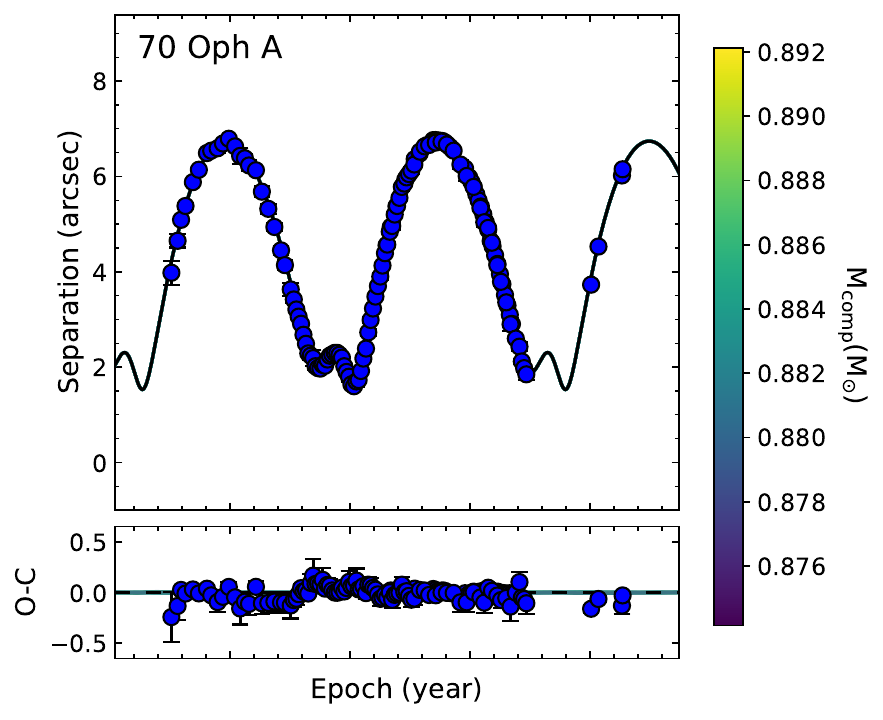}
\includegraphics[width=0.45\linewidth]{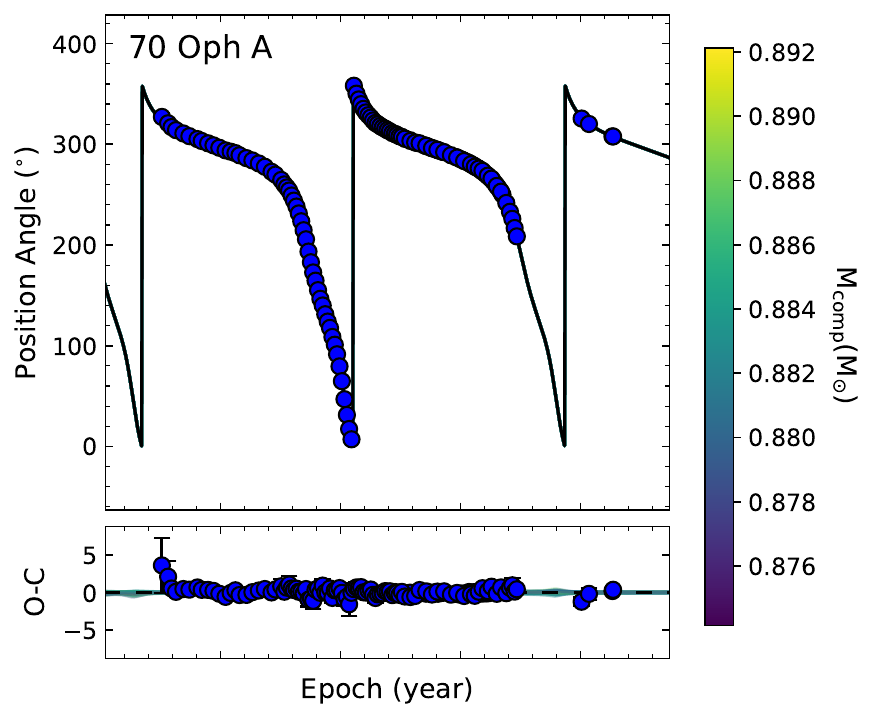}
\caption{Same as Figure~\ref{fig:binary_orbit_2}, but for 70~Oph~A's orbit around 70~Oph~B.}
\label{fig:binary_position_B}
\end{figure*}

\begin{figure*}[!htbp]
\centering
\includegraphics[width=\linewidth]{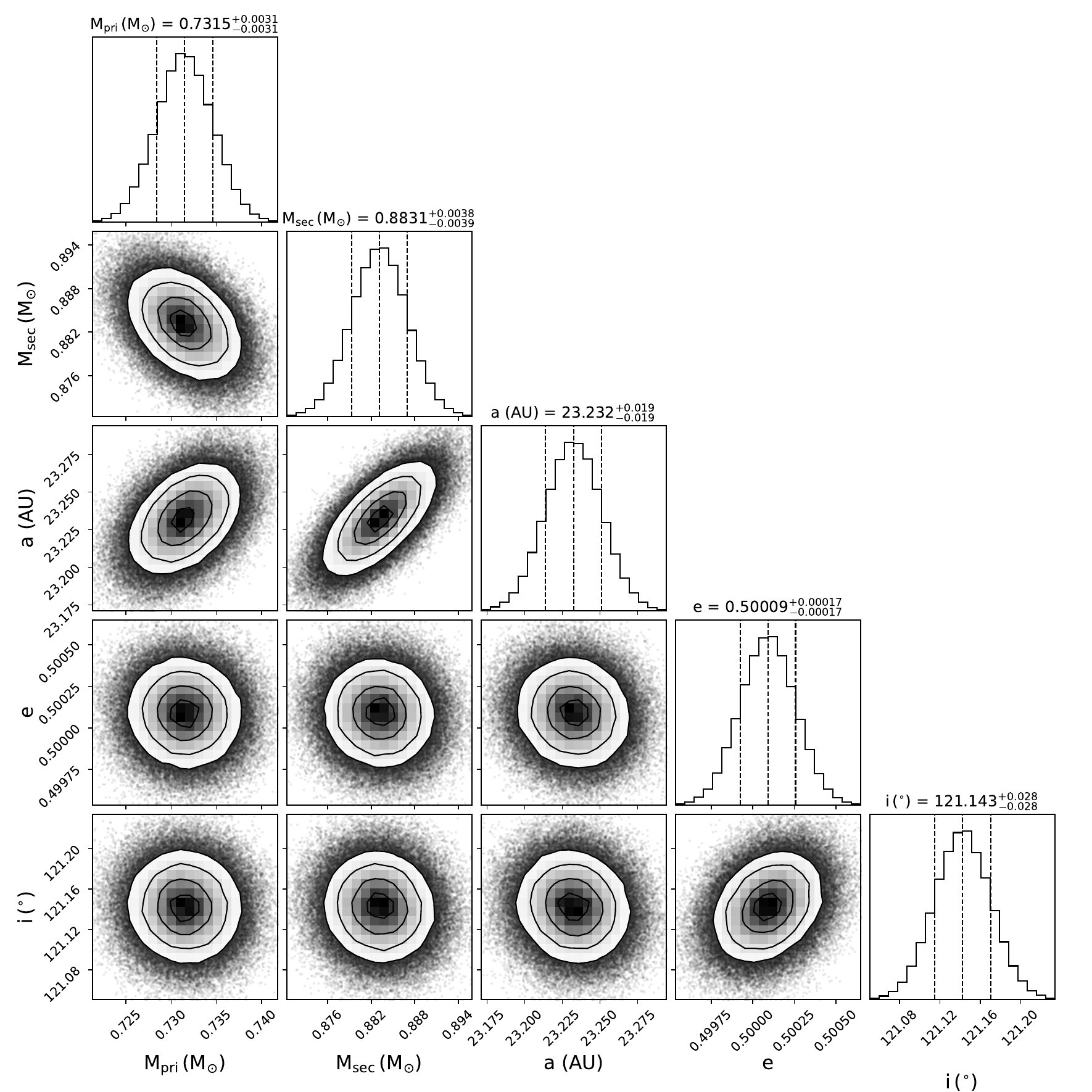}
\caption{
Posterior distributions of key orbital parameters for the independent fit
of 70~Oph~A's orbit around 70~Oph~B (Section~\ref{sec:orbit}), analogous
to Figure~\ref{fig:70OphA_corner} for the primary joint fit. The panels show
marginalized one- and two-dimensional posteriors for the primary and
secondary masses ($M_{\rm pri}$, $M_{\rm sec}$), semimajor axis ($a$),
eccentricity ($e$), and inclination ($i$). Contours represent 1, 2, and
3$\sigma$ credible regions, while vertical dashed lines mark the median and
68\% confidence intervals.}
\label{fig:70OphB_corner}
\end{figure*}

Table~\ref{tab:orbit_results_70OphB} lists the full posterior for the independent orbital fit of 70~Oph~A around 70~Oph~B described in Section~\ref{sec:orbit}. In this fit, 70~Oph~B is treated as the primary (using its own radial velocities and the relative astrometry), with
70~Oph~A as the companion. Gaussian priors on the dynamical masses were adopted from the posterior of the primary joint fit (Table~\ref{tab:orbit_results}). Figures~\ref{fig:binary_orbit_b} and~\ref{fig:binary_position_B} show the resulting astrometric and radial-velocity orbits, and Figure~\ref{fig:70OphB_corner} shows the corresponding posterior distributions for the key orbital parameters.

\begin{deluxetable*}{lccc}
\tablecaption{MCMC Orbit Fitting Results for 70 Ophiuchi A around 70 Ophiuchi B \label{tab:orbit_results_70OphB}}
\tablehead{
\colhead{Parameter} & \colhead{Median $\pm$ 68\% CI} & \colhead{95\% CI} & \colhead{Prior}
}
\startdata
\multicolumn{4}{c}{\textit{Fitted Parameters}} \\
\hline
RV jitter (Mills/Lick) (\ms) & $0.093^{+48}_{-0.093}$ & (0.0, 549.914) & uniform \\
RV jitter (Victoria/DAO) (\ms) & $0.075^{+33}_{-0.075}$ & (0.0, 402.473) & uniform \\
RV jitter (PFS/Magellan) (\ms) & $8.69^{+1.1}_{-0.93}$ & (7.011, 11.203) & uniform \\
\hline
$M_{\rm pri}$ (\msun) & $0.8831^{+0.0038}_{-0.0039}$ & (0.876, 0.891) & Gaussian (70~Oph~A posterior) \\
$M_{\rm sec}$ (\msun) & $0.7315^{+0.0031}_{-0.0031}$ & (0.725, 0.738) & Gaussian (70~Oph~A posterior) \\
Semimajor axis $a$ (au) & $23.232^{+0.019}_{-0.019}$ & (23.196, 23.269) & uniform \\
Semimajor axis $a$ ($\arcsec$) & $4.54339^{+0.00084}_{-0.00084}$ & (4.54174, 4.54505) & -- \\
$\sqrt{e}\,\sin\omega$ & $0.16070^{+0.00071}_{-0.00071}$ & (0.159, 0.162) & uniform \\
$\sqrt{e}\,\cos\omega$ & $0.68867^{+0.00017}_{-0.00017}$ & (0.688, 0.689) & uniform \\
Inclination $i$ ($^\circ$) & $121.143^{+0.028}_{-0.028}$ & (121.088, 121.198) & $\sin i$ \\
PA of ascending node $\Omega$ ($^\circ$) & $121.658^{+0.032}_{-0.032}$ & (121.594, 121.722) & uniform \\
Mean longitude ($^\circ$) & $119.182^{+0.049}_{-0.049}$ & (119.085, 119.278) & uniform \\
Parallax (mas) & $195.56^{+0.15}_{-0.14}$ & (195.275, 195.850) & Gaussian (Gaia EDR3) \\
\hline
\multicolumn{4}{c}{\textit{Derived Parameters}} \\
\hline
Period (yrs) & $88.126^{+0.010}_{-0.010}$ & (88.106, 88.147) & -- \\
Argument of periastron $\omega$ ($^\circ$) & $13.135^{+0.059}_{-0.058}$ & (13.020, 13.251) & -- \\
Eccentricity $e$ & $0.50009^{+0.00017}_{-0.00017}$ & (0.500, 0.500) & -- \\
Semimajor axis (mas) & $4543.39^{+0.84}_{-0.84}$ & (4541.738, 4545.046) & -- \\
Periastron time $T_0$ (JD) & $2477903.9^{+8.4}_{-8.4}$ & (2477887.314, 2477920.501) & -- \\
Mass ratio $q$ ($M_{\rm sec}/M_{\rm pri}$) & $0.8283^{+0.0060}_{-0.0059}$ & (0.8167, 0.8401) & -- \\
\enddata
\tablecomments{Here the mass ratio is computed as $q = M_{\rm sec}/M_{\rm pri}$ to use the same convention as Table~\ref{tab:orbit_results}.}
\end{deluxetable*}

\newpage
\clearpage
\onecolumngrid

\section{Radial Velocity Time Series of 70 Oph A}
\label{app:rvdata}

Table~\ref{tab:rvdata} lists the nightly-binned radial velocity measurements of 70~Oph~A obtained in this work with the Planet Finder Spectrograph (PFS) on the 6.5\,m Magellan Clay Telescope. Each nightly bin is a weighted-average measurement, with weights computed as the inverse
square of the individual RV uncertainties. We provide these data in full below.

\startlongtable
\begin{deluxetable}{cccc}
\tablecaption{PFS Radial Velocity Measurements of 70~Oph~A \label{tab:rvdata}}
\tablehead{
\colhead{BJD$_{\mathrm{TDB}}$} &
\colhead{RV (m s$^{-1}$)} &
\colhead{$\sigma_{\rm RV}$ (m s$^{-1}$)} &
\colhead{Instrument}
}
\startdata
2460153.564 & 11.529 & 0.271 & PFS \\
2460154.553 & 10.998 & 0.232 & PFS \\
2460155.556 & 13.450 & 0.230 & PFS \\
2460156.589 & 11.598 & 0.168 & PFS \\
2460157.589 & 1.364 & 0.173 & PFS \\
2460158.592 & -12.281 & 0.190 & PFS \\
2460159.586 & -25.571 & 0.185 & PFS \\
2460161.623 & -11.867 & 0.208 & PFS \\
2460162.566 & 6.541 & 0.296 & PFS \\
2460181.518 & -8.198 & 0.260 & PFS \\
2460182.538 & 11.226 & 0.263 & PFS \\
2460183.521 & 19.503 & 0.271 & PFS \\
2460184.557 & 13.752 & 0.262 & PFS \\
2460185.526 & 2.567 & 0.271 & PFS \\
2460186.497 & -4.564 & 0.260 & PFS \\
2460191.506 & 17.932 & 0.286 & PFS \\
2460210.505 & 6.931 & 0.244 & PFS \\
2460211.490 & 13.108 & 0.249 & PFS \\
2460212.499 & 18.638 & 0.244 & PFS \\
2460213.490 & 15.480 & 0.259 & PFS \\
2460214.485 & 15.907 & 0.257 & PFS \\
2460215.496 & 13.299 & 0.262 & PFS \\
2460216.489 & 11.534 & 0.249 & PFS \\
2460217.494 & 3.021 & 0.275 & PFS \\
2460218.489 & -9.372 & 0.257 & PFS \\
2460485.658 & -13.142 & 0.304 & PFS \\
2460486.667 & -14.816 & 0.211 & PFS \\
2460487.676 & -14.952 & 0.212 & PFS \\
2460488.698 & -12.636 & 0.203 & PFS \\
2460490.683 & -1.428 & 0.204 & PFS \\
2460491.660 & 9.010 & 0.209 & PFS \\
2460492.674 & 16.078 & 0.210 & PFS \\
2460509.634 & -2.785 & 0.224 & PFS \\
2460510.619 & 1.533 & 0.224 & PFS \\
2460513.587 & 15.351 & 0.308 & PFS \\
2460514.627 & 7.410 & 0.261 & PFS \\
2460515.631 & -5.070 & 0.221 & PFS \\
2460516.601 & -14.288 & 0.274 & PFS \\
2460545.582 & -21.430 & 0.326 & PFS \\
2460546.567 & -15.188 & 0.330 & PFS \\
2460547.472 & -8.259 & 0.302 & PFS \\
2460548.471 & -4.488 & 0.297 & PFS \\
2460598.498 & -27.236 & 0.268 & PFS \\
2460599.509 & -26.955 & 0.264 & PFS \\
2460603.511 & -20.706 & 0.288 & PFS \\
2460747.924 & -4.862 & 0.344 & PFS \\
2460748.900 & -11.086 & 0.338 & PFS \\
2460753.897 & -5.038 & 0.276 & PFS \\
\enddata
\end{deluxetable}

\FloatBarrier

\section{Radial Velocity Time Series of 70 Oph B}
\label{app:rvdata_B}

Table~\ref{tab:rvdata_B} lists the nightly-binned radial velocity
measurements of 70~Oph~B obtained in this work with the Planet Finder
Spectrograph (PFS) on the 6.5\,m Magellan Clay Telescope. Each nightly bin is a weighted-average measurement, with weights computed as the inverse
square of the individual RV uncertainties.

\startlongtable
\begin{deluxetable}{cccc}
\tablecaption{PFS Radial Velocity Measurements of 70~Oph~B \label{tab:rvdata_B}}
\tablehead{
\colhead{BJD$_{\mathrm{TDB}}$} &
\colhead{RV (m s$^{-1}$)} &
\colhead{$\sigma_{\rm RV}$ (m s$^{-1}$)} &
\colhead{Instrument}
}
\startdata
2460153.573 & -7.925 & 0.304 & PFS \\
2460154.563 & -7.636 & 0.241 & PFS \\
2460155.586 & -8.427 & 0.236 & PFS \\
2460156.606 & -9.973 & 0.184 & PFS \\
2460157.600 & -11.880 & 0.174 & PFS \\
2460158.604 & -10.017 & 0.202 & PFS \\
2460159.592 & -9.584 & 0.193 & PFS \\
2460160.528 & -10.340 & 0.266 & PFS \\
2460161.632 & -7.669 & 0.218 & PFS \\
2460162.578 & -6.553 & 0.411 & PFS \\
2460181.528 & -6.386 & 0.283 & PFS \\
2460182.548 & -5.152 & 0.297 & PFS \\
2460183.531 & -7.572 & 0.315 & PFS \\
2460184.567 & -9.720 & 0.343 & PFS \\
2460185.536 & -5.615 & 0.328 & PFS \\
2460186.507 & 3.785 & 0.360 & PFS \\
2460191.517 & -4.094 & 0.336 & PFS \\
2460211.500 & -5.300 & 0.278 & PFS \\
2460213.499 & -5.899 & 0.305 & PFS \\
2460214.495 & -11.433 & 0.298 & PFS \\
2460215.508 & -11.558 & 0.310 & PFS \\
2460216.499 & -8.853 & 0.309 & PFS \\
2460217.504 & -2.245 & 0.329 & PFS \\
2460218.499 & -0.118 & 0.318 & PFS \\
2460485.669 & 13.019 & 0.293 & PFS \\
2460486.678 & 7.956 & 0.223 & PFS \\
2460487.687 & 3.170 & 0.247 & PFS \\
2460488.708 & 0.915 & 0.199 & PFS \\
2460490.690 & 24.325 & 0.190 & PFS \\
2460491.670 & 30.103 & 0.192 & PFS \\
2460492.683 & 20.842 & 0.185 & PFS \\
2460509.643 & 23.053 & 0.207 & PFS \\
2460510.629 & 29.739 & 0.217 & PFS \\
2460513.597 & 6.234 & 0.341 & PFS \\
2460514.595 & 23.301 & 0.485 & PFS \\
2460515.646 & 1.445 & 0.249 & PFS \\
2460516.611 & -0.505 & 0.322 & PFS \\
2460545.622 & 13.504 & 0.368 & PFS \\
2460546.578 & 23.722 & 0.502 & PFS \\
2460547.486 & 20.057 & 0.293 & PFS \\
2460747.930 & 24.306 & 0.390 & PFS \\
2460748.894 & 25.139 & 0.430 & PFS \\
2460752.895 & 23.294 & 0.329 & PFS \\
2460753.887 & 19.957 & 0.284 & PFS \\
\enddata
\end{deluxetable}

\end{document}